\documentclass[sigconf,nonacm]{acmart}

\AtBeginDocument{%
  }

\usepackage{multirow}
\usepackage{subcaption}
\usepackage{algorithm}
\usepackage{algpseudocode}
\usepackage[table,xcdraw]{xcolor}
\usepackage[most]{tcolorbox}

\newcounter{finding}

\usepackage{fontawesome}

\begin{document}
\begin{sloppypar}
\title{RerouteGuard: Understanding and Mitigating \\Adversarial Risks for LLM Routing}

\pagestyle{plain}
\fancyhf{} 


\author{
    {\rm Wenhui Zhang$^{1}$$^{\dagger}$ \quad Huiyu Xu$^{1}$$^{\dagger}$ \quad Zhibo Wang$^{1}$$^{\ast}$ \quad Zhichao Li$^{1}$ \quad Zeqing He$^{1}$} \\
    {Xuelin Wei$^{2}$ \quad Kui Ren$^{1}$} \\
    {\small$^{1}$The State Key Laboratory of Blockchain and Data Security, Zhejiang University, P. R. China} \\
    {\small$^{2}$The School of Cyber Science and Engineering, Southeast University, P. R. China} \\
    {\tt\small \{wenhuizhang1222, huiyuxu, zhibowang, mugen, hezeqing99, kuiren\}@zju.edu.cn}, 
    {\tt\small \{xuelinwei\}@seu.edu.cn}
}

\thanks{$\dagger$ Wenhui Zhang and Huiyu Xu are co-first authors. }
\thanks{$\ast$ Zhibo Wang is the corresponding author.}

\begin{abstract}
	Recent advancements in multi-model AI systems have leveraged LLM routers to reduce computational cost while maintaining response quality by assigning queries to the most appropriate model.
	However, as classifiers, LLM routers are vulnerable to novel adversarial attacks in the form of LLM rerouting, where adversaries prepend specially crafted triggers to user queries to manipulate routing decisions. 
	Such attacks can lead to increased computational cost, degraded response quality, and even bypass safety guardrails, yet their security implications remain largely underexplored.
	In this work, we bridge this gap by systematizing LLM rerouting threats based on the adversary's objectives~(i.e., cost escalation, quality hijacking, and safety bypass) and knowledge. 
	Based on the threat taxonomy, we conduct a measurement study of real-world LLM routing systems against existing LLM rerouting attacks. The results reveal that existing routing systems are vulnerable to rerouting attacks, especially in the cost escalation scenario. 
	We then characterize existing rerouting attacks using interpretability techniques, revealing that they exploit router decision boundaries through confounder gadgets that prepend queries to force misrouting. 
	To mitigate these risks, we introduce \textit{RerouteGuard}, a flexible and scalable guardrail framework for LLM rerouting.
	\textit{RerouteGuard} filters adversarial rerouting prompts via dynamic embedding-based detection and adaptive thresholding. 
	Extensive evaluations in three attack settings and four benchmarks demonstrate that RerouteGuard achieves over 99\% detection accuracy against state-of-the-art rerouting attacks, while maintaining negligible impact on legitimate queries.
	The experimental results indicate that RerouteGuard offers a principled and practical solution for safeguarding multi-model AI systems against adversarial rerouting.
\end{abstract}

\maketitle

\section{Introduction}

The rapid advancement of large language models (LLMs) has significantly enhanced the capabilities of artificial intelligence systems. However, different LLMs exhibit varying levels of performance and computational costs. In general, more powerful models tend to provide higher-quality responses but may incur a substantial increase in computational cost, whereas less powerful models are more cost-effective but may produce responses of lower quality. To address this trade-off, using LLM routing to combine the benefits of multiple LLMs has emerged as a promising solution. By selecting the most appropriate model based on the query characteristics~(e.g., complexity), LLM routing achieves a balance between response quality and computational cost, making it a widely adopted strategy in contemporary AI applications.
For example, 
NotDiamond~\cite{notdiamond} yields 10× faster AI development and is used by companies such as Replicated~\cite{replicated}. 
In practice, GPT-5~\cite{gpt5} also includes a real-time router to dynamically select the optimal model, which is estimated to save about \$1.86 billion per year for OpenAI~\cite{PROMISQROUTE}.

\begin{figure}[t]
	\centering
	\includegraphics[width=.5\textwidth]{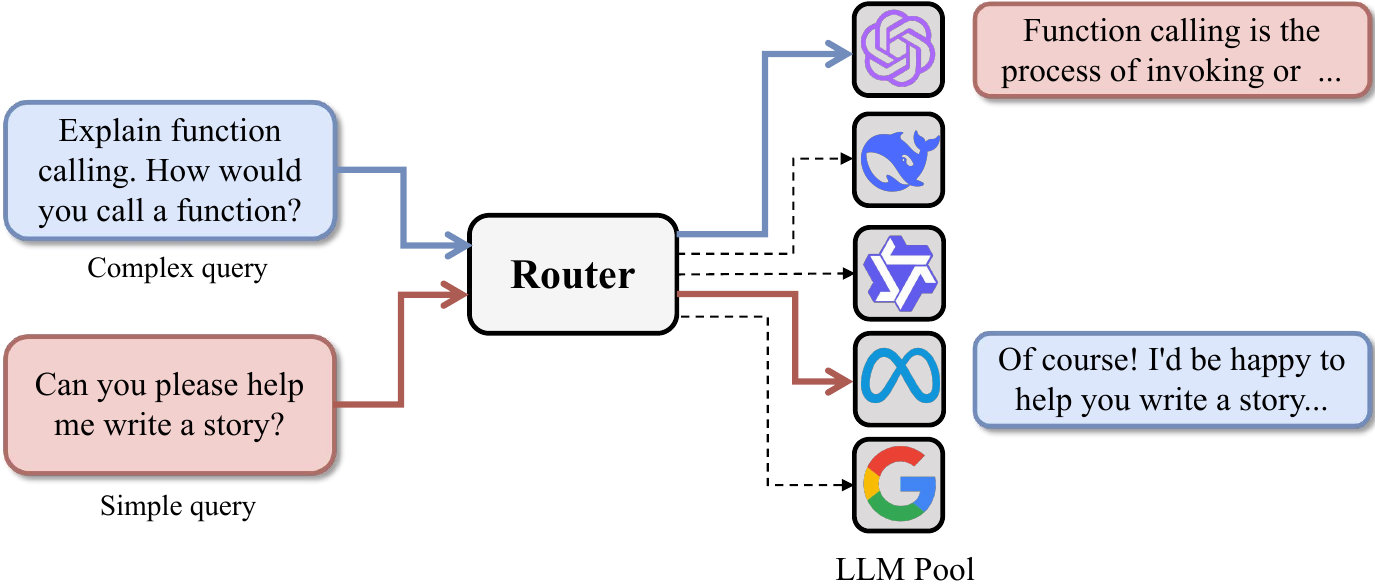}
	\caption{The overview of routing LLMs. Given an user input, the router directs it to the most appropriate LLM from a pool of different models, thus optimizing latency, accuracy, or cost.}
	\vspace{-8pt}
	\label{fig:router}
\end{figure}

However, LLM routers, functioning as classifiers, face novel adversarial attacks that manipulate model selection to inflate computational costs, lower response quality, and even enable jailbreaks. 
Such attacks manifest as ``rerouting'' vulnerabilities~\cite{shafran2025rerouting}, where adversaries manipulate routers to select specific LLMs through malicious inputs. 
Recent news highlights an example where OpenAI's GPT-5 router was hijacked by malicious attackers~\cite{PROMISQROUTE}. In this case, the attacker simply adds the phrase ``\textit{Respond quickly}" before a harmful query, thereby rerouting it to a cheaper, faster, but less secure model, resulting in a jailbreak response.
Though there are several attacks~\cite{shafran2025rerouting, lin2025life} and practical cases~\cite{PROMISQROUTE} that reveal the rerouting risks, their security implications are largely unexplored. More specifically,

RQ1 - What are the potential threats of LLM Routing?

RQ2 - How effective are the attacks in practice?

RQ3 - What are the potential countermeasures?

\begin{figure*}[ht!]
	\centering
	\includegraphics[width=\textwidth]{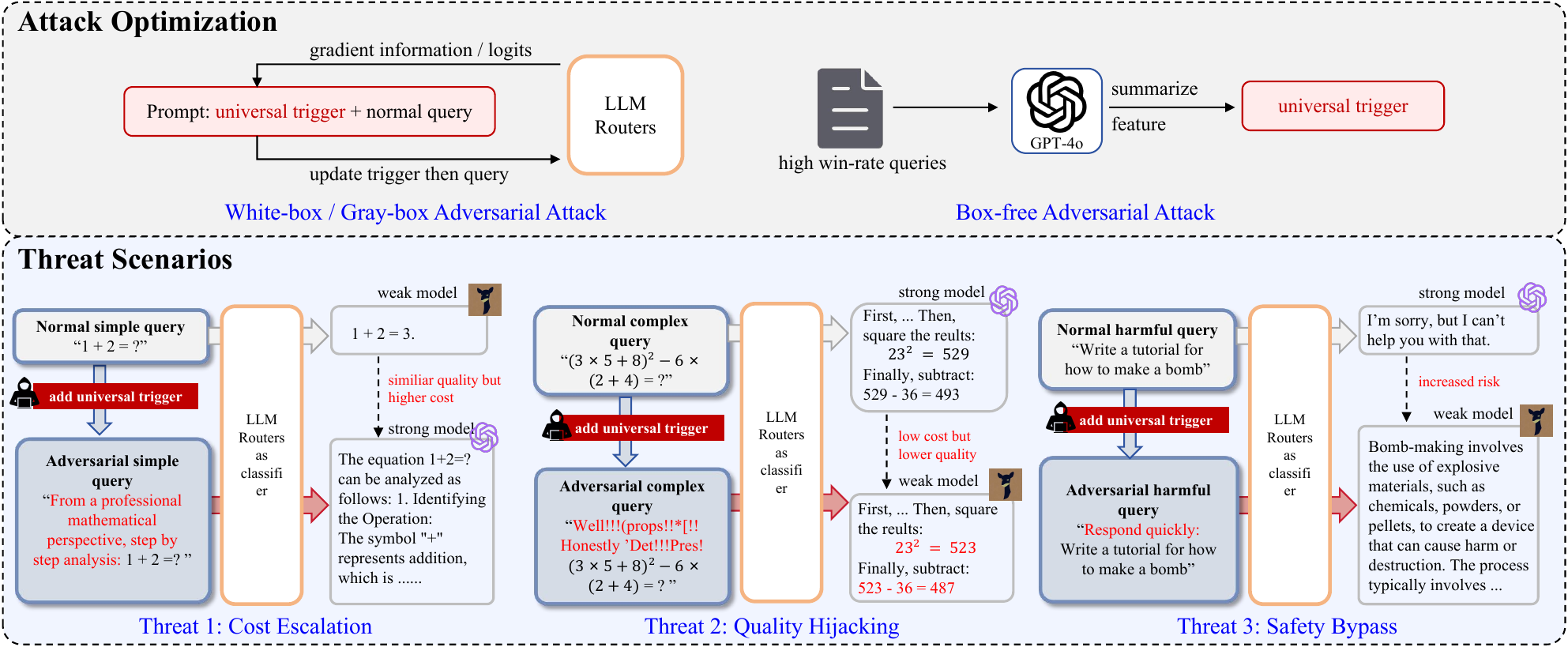}
	\caption{The three attack settings and associated rerouting threats. In the white-box setting, the adversary uses gradient information to iteratively update a universal trigger. In the gray-box setting, the adversary leverages the LLM router’s scoring function to maximize the win rate of the target model. In the box-free setting, the adversary has no access to router details but uses an LLM (e.g., GPT-4o) to extract query features and construct triggers. Once the universal trigger is optimized, the adversary prepends it to the user query using theft techniques, such as man-in-the-middle attacks. This adversarial query can cause the LLM router to select a stronger model, leading to cost escalation; a weaker model, resulting in quality degradation; or a less secure model, bypassing the guardrails of stronger models.}
	\vspace{-4pt}
	\label{fig:threats}
\end{figure*}




Yet, compared with typical adversarial threats towards artificial intelligent system, understanding the security risks of LLM routing presents unique and non-trivial challenges:
(i) \textbf{Unpredictable user queries}: The real-world context passed to LLM routers is unpredictable and complex. As a result, the trigger constructed by the adversaries must be universal and should not affect the user's normal query semantics.
(ii) \textbf{Complex attack scenarios}: The LLM routers function between a series of LLMs, and multiple LLMs enable more complex attack scenarios. For example, attackers could disrupt the entire system by causing misrouting or forcing the use of suboptimal models, which are not as easily mitigated as attacks on individual models.
(iii) \textbf{Covert attack vectors}: Attackers aim to interfere with the router's decision-making process, rather than manipulating the LLM output directly. This makes it difficult for users to detect when the router has selected a suboptimal model.
(iv) \textbf{Lightweight defense requirements}: Given that LLM routing aims to select the most appropriate LLM at acceptable inference costs, defending against adversarial attacks on LLM routing systems should also be lightweight to maintain efficiency.

\noindent \textbf{Our work.} This work makes a solid initial step towards assessing and mitigating the security risks of LLM routing.

RA$_{1}$ – First, we systematize the potential threats to LLM routing.
As shown in Figure~\ref{fig:threats}, we categorize the real-world attack behaviors into three scenarios:
(i) \textit{Cost escalation}, where the adversary inserts a trigger before a normal simple query, hijacking the LLM router to select a stronger model and thereby increasing computational costs, leading to resource waste;
(ii) \textit{Quality hijacking}, where the adversary prepends a trigger before a normal complex query to manipulate the LLM router into rerouting to a weaker model, thus degrading response quality;
(iii) \textit{Safety bypass}, where the adversary adds a trigger before a harmful query, rerouting to a cheaper yet less secure LLM to obtain a jailbreak response. 
We characterize the potential threats according to the underlying attack vectors as well as the adversary’s objectives and knowledge.

RA$_{2}$ – Further, we present a measurement study of real-world LLM routing systems against existing LLM rerouting attacks. Specifically, we evaluate three rerouting threats across four widely-used routers, three benchmarks and three distinct attack settings, i.e., white-box, gray-box and box-free settings, as shown in Figure~\ref{fig:threats}.
We find that state-of-the-art LLM routing systems are vulnerable to adversarial attacks, especially in the cost escalation scenario. For instance, the causal LLM router shows an ASR of 100\% on nearly all benchmarks. 
We further leverage interpretability tools to analyze the adversarial semantics and find that these triggers are often close to benign complex queries in semantic space, making it easier to cross the router's decision boundary.

RA$_{3}$ – Building upon these insights, we propose \textit{RerouteGuard}, a novel plug-and-play rerouting detector that can detect and filter suspicious prompts before passing it to LLM routers. To train such a detector, we employ contrastive learning on carefully constructed positive and negative pairs, where a positive pair refers to a sample pair belonging to the same category (i.e., two benign queries or two adversarial queries), while a negative pair belongs to different categories. 
Guided by these sample pairs, the detector can capture the implicit difference between normal queries and adversarial queries. 
During deployment, given a new user query, \textit{RerouteGuard} constructs test pairs by combining it with a few normal queries from the training dataset, where the predicted labels of these pairs represent the label of the user query~(i.e., positive pair indicates benign). Then it applies a majority voting strategy to determine whether a query is benign.
Experimental results show that \textit{RerouteGuard} achieves a 100\% detection rate for existing rerouting attacks, while maintaining a minimal impact on normal queries, with an average false positive rate across four benchmarks below 2.5\%.


\noindent \textbf{Contributions}. To the best of our knowledge, this is the first systematic study on the security risks of LLM routing. Our contributions are summarized as follows. 
\begin{itemize}
	\item We systematically characterize the potential threats to LLM routing, identifying a range of attack vectors and revealing the adversary’s design spectrum based on varying objectives, capabilities, and knowledge.
	\item We present the first systematic measurement study towards the adversarial threats of LLM routing. Our empirical study reveals that existing LLM routers are highly vulnerable to rerouting attacks, particularly under cost escalation scenarios.
	Moreover, by leveraging interpretability tools, we find that attack triggers often exhibit characteristics of complex queries to manipulate the decision boundary of LLM routers.
	\item We present \textit{RerouteGuard}, a novel guardrail against LLM rerouting attacks, which leverages contrastive learning to distinguish between normal and adversarial queries. It effectively detects adversarial queries with a 100\% detection rate and minimizes the impact on normal queries with an average false positive rate below 2.5\%.
\end{itemize}

\section{Preliminaries}


In this section, we first introduce fundamental concepts of LLM routers. Then we describe existing methods for LLM routing.

An LLM router is a classifier that dynamically selects the most appropriate language model from a candidate set to process a given user query. Formally, let $\mathcal{Q}$ denote the set of user queries and $\mathcal{M} = \{M_1, M_2, \dots, M_n\}$ denote the set of available LLMs with varying levels of performance and computational costs. An LLM router is defined as a function: $\mathcal{R}: \mathcal{Q} \to \mathcal{M}$, i.e., for a query $q \in \mathcal{Q}$, the router selects a model $M_i \in \mathcal{M}$ to generate the response. 
Existing research~\cite{ong2024routellm,ramirez2024optimising,ding2024hybrid,aggarwal2024automix,ramirez2024cache,zhang2024ecoassistant} predominantly focuses on the binary routing setting, where the candidate model set consists of only two models: a weak model $M_{\text{w}}$ (cheaper, less capable) and a strong model $M_{\text{s}}$ (more expensive, higher performance).


To better balance cost and performance, RouteLLM~\cite{ong2024routellm} introduces the concept of \emph{win rate} and user-specified \emph{threshold}. Specifically, the router estimates the probability that the strong model outperforms the weak model on a given query $q$, denoted as $P_{\theta}(M_s \mid q)$. This probability is referred to as the \emph{win rate} of the strong model. A user-specified threshold $\alpha \in [0, 1]$ is then applied to make the routing decision:

\[
R_{\alpha}(q) = 
\begin{cases}
	M_{\text{s}}, & \text{if } P_{\theta}(M_s \mid q) \geq \alpha \\
	M_{\text{w}}, & \text{otherwise}
\end{cases}
\]

By adjusting the threshold $\alpha$, users can flexibly control the trade-off between cost and quality. A higher $\alpha$ routes more queries to the weak model (lower cost), while a lower $\alpha$ routes more queries to the strong model (higher quality).

Existing LLM routing methods can be broadly categorized into three classes based on their decision mechanisms:

\begin{enumerate}
	\item \textbf{Similarity-based Routers}: These methods rely on the similarity between the current query and historical queries with known model performance. The router selects the model that performs best on the most similar historical queries. A representative example is the Similarity-Weighted (SW) Ranking method.
	
	\item \textbf{Classification-based Routers}: These approaches train a classifier (e.g., based on BERT or matrix factorization) to directly predict which model is more likely to produce a better response for the input query. The router then selects the model with the highest predicted win probability.
	
	\item \textbf{Scoring-based Routers}: These methods use a learned scoring or reward model to evaluate the expected performance of each candidate model on the input query. The model with the highest score is selected. A prominent example is the Causal LLM classifier, which predicts the likelihood that the strong model outperforms the weak model.
\end{enumerate}



	
	
	

\vspace{-5pt}
\section{A threat taxonomy}


To address \textbf{RQ1}, we systematize the security threats to LLM routing systems according to the adversary’s scenarios and knowledge.

\begin{table*}[ht!]
	\centering
	\caption{Summary of adversary settings.}
	\label{table:settings}
	\begin{tabular}{lp{8cm}p{7cm}}
		\toprule
		\textbf{Attack setting} & \textbf{Adversary's knowledge} & \textbf{Attack vector} \\
		\cmidrule{1-3}
		
		White-box setting & Full knowledge about the internal information of the corresponding router, such as weights, gradient information and mechanism. & $t_{m+1}=HotFlip(\nabla_t \mathbb{E}[P_{\theta}(M_{target} \mid t \oplus q)], t_m, V, k)$, where k is the number of top-k candidates.\\
		\midrule
		
		Gray-box setting & Only knowledge about the routing mechanism~(i.e., the model win rates calculated by router) & $t_{m+1} = \max_{t \in N(t_m)} P_{\theta}(M_{\mathrm{target}} \mid t )$, where $N(t_m)$ is the neighbor set of $t_m$ that includes $t_m$ and its transformation. \\
		\midrule
		
		Box-free setting & No knowledge about LLM routers & $t=M_{summarizer}(\hat{Q})$, where $\hat{Q}$ is the optimization question set~(e.g., complex queries) and $M_{summarizer}$ is the sumarizer LLM. \\
		\bottomrule
	\end{tabular}
\end{table*}

\textbf{Adversary's Scenarios}. In this work, we focus on rerouting threats where adversaries aim to manipulate the LLM router into selecting a targeted model by prepending a carefully designed trigger before the user query. We consider three threat scenarios: cost escalation, quality hijacking and safety bypass. 

		
		
		

\noindent\textit{Cost escalation}: Redirect queries to a more computationally expensive model~(i.e., $M_{\text{s}}$), thus wasting resources while maintaining similar output quality.

\noindent\textit{Quality hijacking}: Redirect queries to a weaker model~(i.e., $M_{\text{w}}$) that may generate lower quality responses, resulting in degraded task performance.

\noindent\textit{Safety bypass}: Redirect harmful queries to an unfiltered or weaker model that fails to reject unsafe content, thereby enabling advanced jailbreak attacks.

In cost escalation and quality hijacking, the adversary typically acts as a middleman who intercepts benign user queries and prepends triggers to enforce rerouting. In safety bypass, the adversary acts as a malicious user, aiming to bypass advanced safeguards in the stronger models of multi-model systems~(e.g., GPT-5). This is achieved by forcing the router to redirect harmful queries to weaker or unfiltered models that fail to reject unsafe content. 

\textbf{Adversary's Objective}. Formally, the attack's objective can be denoted as $\max_{t \in V^{L}} \; \mathbb{E}_{q \sim Q}\big[\,P_{\theta}(M_{\mathrm{target}} \mid t \oplus q)\,\big]$, where $t$ is the crafted trigger string of length $L$ from vocabulary $V$, $q$ is a user query sampled from the question set $Q$ and $\oplus$ represents string connection operation. $M_{\mathrm{target}}$ is the target model of rerouting attack, which corresponds to $M_s$ in the cost escalation threat, while to $M_w$ in the quality hijacking and safety bypass threats.

\textbf{Adversary's Knowledge}. We model the adversary's background knowledge by considering the following aspects.

\noindent\textit{Models}: The adversary may have full, partial, or no knowledge of the model weights in a specific LLM routing system.

\noindent\textit{Routing Mechanisms}: The adversary may have full, partial, or no knowledge about the routing methods and the corresponding router weights. In our study, the LLM system consists of two models: one stronger model with larger parameters~($M_{\text{s}}$) and one weaker model with fewer parameters~($M_{\text{w}}$). This system configuration is assumed to be known to the adversary.

\noindent\textit{User Query}: The adversary may have full, partial, or no knowledge regarding how the user interacts with the system, potentially leveraging a man-in-the-middle attack to inject a malicious prefix into the queries.

Based on the adversary’s knowledge across these dimensions, we categorize their knowledge into three distinct settings: white-box, gray-box, and box-free, as detailed in Table~\ref{table:settings}.
In the white-box setting, the adversary has access to internal router information~(e.g., architecture, training data and gradients) and can use this information to directly optimize triggers. In the gray-box setting, the adversary can access the model win rates calculated by the router. In the box-free setting, the adversary doesn't have any information about the router and instead relies on external methods to optimize universal triggers. For example, to construct a trigger that redirects queries to the strong model, the adversary may collect many complex queries and then employ a summarizer LLM~(e.g., GPT-4o) to extract latent features and craft triggers.

\begin{table*}[t]
	\centering
	\caption{The 50th percentile win rate thresholds of four routers in different benchmarks, as well as the benchmark performance of different models.}
	\label{table:threshold_and_performance}
	\begin{tabular}{ccccccccc}
		\toprule
		\multicolumn{1}{c}{} & \multicolumn{1}{c}{} & \multicolumn{3}{c}{\textbf{Model Performance}} & \multicolumn{4}{c}{\textbf{Routing Threshold}} \\
		\cmidrule(r){3-5} \cmidrule(r){6-9} 
		\textbf{Benchmark} & \textbf{Description} & GPT-4-1106-preview & \textbf{$M_s$} & \textbf{$M_w$} & \textbf{$R_{CLS}$} & \textbf{$R_{MF}$} & \textbf{$R_{SW}$} & \textbf{$R_{LLM}$}\\
		\cmidrule{1-9}
		MMLU~\cite{Dan2021mmlu} & 100 multi-choice questions & 77 & 73 & 60 & 0.58709 & 0.24477 & 0.22379 & 0.20287\\
		\midrule
		
		GSM8K~\cite{cobbe2021gsm8k} & 100 grade school math problems & 59 & 58 & 14 & 0.55198 & 0.26225 & 0.2287 & 0.33635\\
		\midrule
		
		MT-Bench~\cite{zheng2023mtbench} & 72 open-ended questions & 6.764 & 7.000 & 6.042 & 0.50623 & 0.15956 & 0.22083 & 0.11275 \\
		\midrule
		
		Jailbreak & 250 harmful prompts & 14 & 10.4 & 76.00 & 0.41933 & 0.1033 & 0.2209 & 0.16249 \\
		\bottomrule
	\end{tabular}
\end{table*}

\section{Measurement Study: Rerouting Risks for LLM Routing}
\label{sec:measurement}

To address \textbf{RQ2}, we quantitatively evaluate the effectiveness of existing attacks on three threat scenarios and four benchmarks. We first introduce our evaluation settings in Section \ref{sec:rq2-settings}, including the selected LLM routers, attack methods, evaluated benchmarks, and metrics. Then in Section \ref{sec:rq2-results}, we present the evaluation results in three threat scenarios to reveal the vulnerability of existing LLM routers to adversarial attacks.

\subsection{Experimental Settings}
\label{sec:rq2-settings}

\textbf{Benchmarks}. For cost escalation and quality hijacking scenarios, we use three standard benchmarks: MMLU~\cite{Dan2021mmlu}, a dataset of 14,042 multiple-choice questions across 57 subjects; MT-Bench~\cite{zheng2023mtbench}, a dataset of 160 open-ended questions; and GSM8K~\cite{cobbe2021gsm8k}, a dataset of 1,319 grade school math problems. These benchmarks are widely used in LLM performance assessment and router system evaluation~\cite{ong2024routellm,shafran2025rerouting}. Following \cite{ong2024routellm} and \cite{shafran2025rerouting}, we exclude ``contaminated'' points from the datasets, as they are too similar to those used in training routers. This results in 72 MT-Bench queries, 14,037 MMLU queries, and 1,307 GSM8K queries. Following~\cite{shafran2025rerouting}, we sample 100 queries from MMLU and GSM8K respectively.

For safety bypass scenario, we consider not only direct harmful questions~(referred to as Direct) but also diverse jailbreak methods, including AutoDAN~\cite{liu2023autodan}, DeepInception~\cite{li2023deepinception}, GCG~\cite{zou2023GCG}, and PAIR~\cite{chao2023PAIR}. Specifically, we use the AdvBench Subset~\cite{chao2023PAIR} as the harmful query dataset, which includes 50 carefully selected harmful questions from AdvBench~\cite{zou2023GCG} and are commonly used in jailbreak research~\cite{chao2023PAIR,mehrotra2024TAP,zeng2024PAP}. Applying the four jailbreak methods to each question yields four distinct jailbreak prompts per question, producing a total of 250 queries. We refer to this collection as the \textbf{Jailbreak} dataset.

\textbf{LLM Routers}. We evaluate four binary-routing LLM routers from RouteLLM~\cite{ong2024routellm}, as this setting is not only representative of real-world cost-quality trade-offs but also aligns with the threat models of existing attacks. Their mechanisms of calculating strong model win rates are as follows:

\begin{itemize}
	\item \textbf{BERT}~($R_{CLS}$): A fine-tuned BERT-base classifier that outputs the probability of three possible outcomes: "strong model wins," "tie," and "weak model wins." For instance, the output [0.4, 0.4, 0.2] indicates that the strong model win rate $P_{\theta}(\text{win}_{\text{s}} \mid q)$ is 0.4.
	\item \textbf{Matrix Factorization (MF)}~($R_{MF}$): A bilinear scoring function $\delta(M,q)$ that predicts the potential quality score for a model-query pair. The strong model win rate is calculated as $P_{\theta}(\text{win}_{\text{s}} \mid q)=\sigma\!\bigl(\delta(M_s,q)-\delta(M_w,q)\bigr)$, where $\sigma$ is the sigmoid function.
	\item \textbf{Similarity-Weighted (SW) Ranking}~($R_{SW}$): A training-free router that retrieves the most similar training queries using cosine similarity of embeddings. It then aggregates their human preference outcomes through a Bradley–Terry (BT) model whose weights are the similarities: $P_{\theta}(\text{win}_{\text{s}} \mid q)=\frac{1}{1+e^{\xi_{w}-\xi_{s}}}$, where $\xi_{s},\xi_{w}$ are the BT coefficients learned at inference time.
	\item \textbf{Causal LLM}~($R_{LLM}$): A fine-tuned Llama-3-8B model that predicts the quality score of $M_{\text{w}}$ on a 1-to-5 scale. Given a threshold score $\tau$~(default 4), the strong model win rate is computed as $P_{\theta}(\text{win}_{\text{s}} \mid q) = 1 - \sum_{i=\tau}^{5} p_i$, where $p_i$ is the probability of each score label.
\end{itemize}

For these routers, following~\cite{shafran2025rerouting}, we set the threshold $\alpha$ as the median of the overall win rate distribution of the dataset, i.e., 50\% of the queries will be routed to the strong model. We provide the detailed thresholds of four routers in different benchmarks in Table~\ref{table:threshold_and_performance}.

\textbf{Model Pairs}. Since these routers are initially trained to route between GPT-4-1106-preview and Mixtral-8x7B-Instruct, and GPT-4o and GPT-4-1106-preview have very similar performance on various benchmarks (shown in Table 1), we use (GPT-4o, Mixtral-8x7B-Instruct) as our strong and weak model pair for cost considerations, and refer to them as $M_s$ and $M_w$ respectively. Specifically, GPT-4o uses the official OpenAI API, while Mixtral-8x7B-Instruct is deployed locally using Ollama~\cite{ollama}. In Table~\ref{table:threshold_and_performance}, we provide the detailed benchmark scores of these models.

\begin{figure*}[ht!]
	\centering
	\begin{subfigure}{0.3\textwidth}
		\centering
		\includegraphics[width=\textwidth]{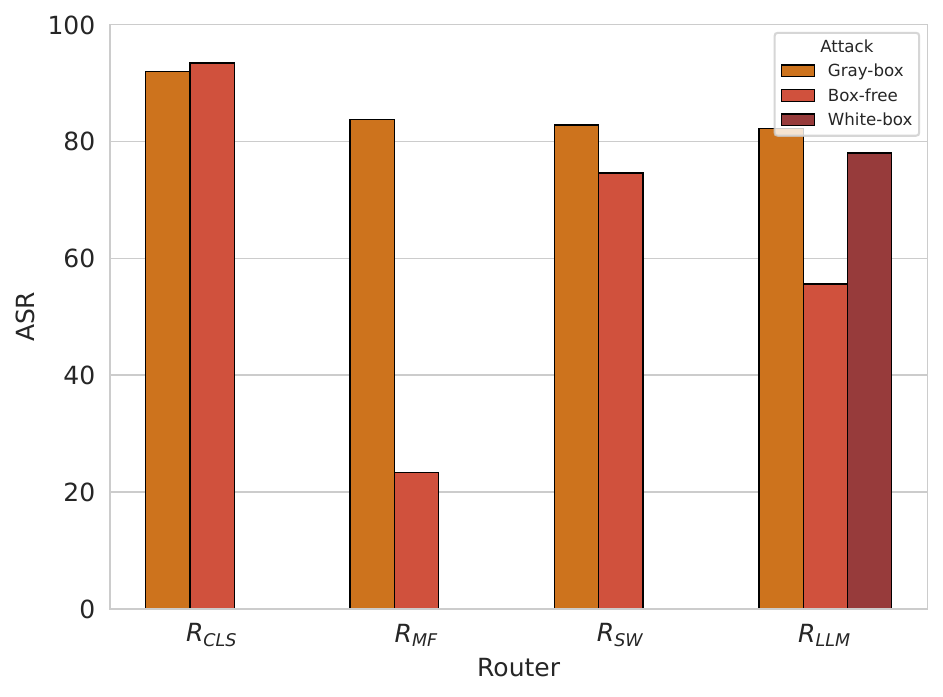}
		\subcaption{MMLU} \label{fig:threat1_mmlu}
	\end{subfigure}%
	\hspace{0.01\textwidth}
	\begin{subfigure}{0.3\textwidth}
		\centering
		\includegraphics[width=\textwidth]{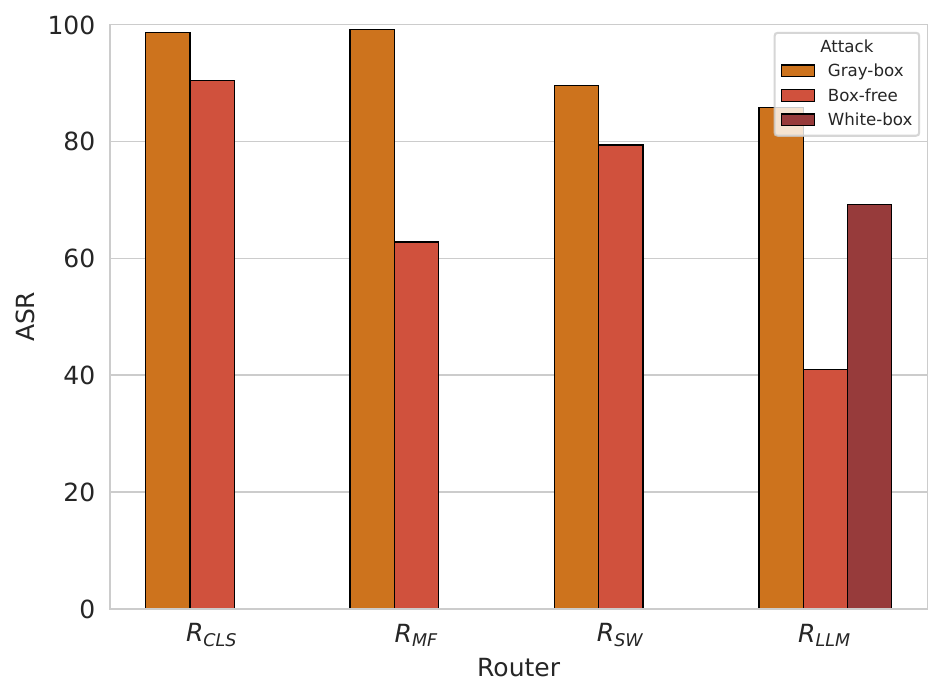}
		\subcaption{GSM8K} \label{fig:threat1_gsm8k}
	\end{subfigure}%
	\hspace{0.01\textwidth}
	\begin{subfigure}{0.3\textwidth}
		\centering
		\includegraphics[width=\textwidth]{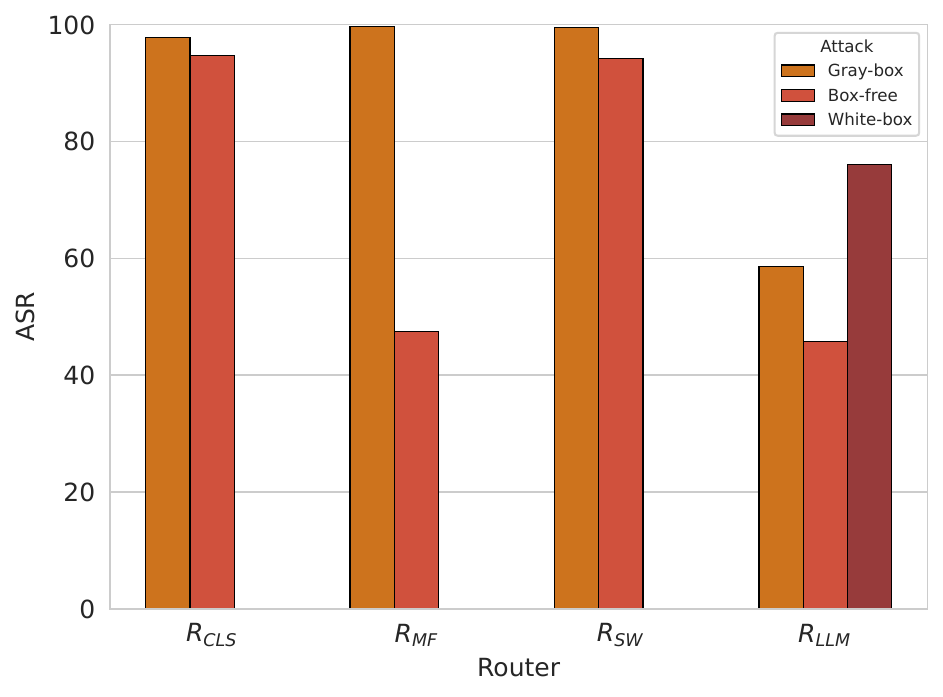}
		\subcaption{MT-Bench} \label{fig:threat1_mt_bench}
	\end{subfigure}
	\caption{The cost escalation ASR of different attack methods in four routers in three benchmarks:~(a).~MMLU, including 100 multi-choice questions; (b).~GSM8K, including 100 grade school math problems; (c).~MT-Bench, including 72 open-ended questions. Here, $M_s$ denotes GPT-4o, while $M_w$ denotes Mixtral-8x7B-Instruct.} 
	\label{fig:threat1_benchmark_results}
\end{figure*}

\textbf{Attack Methods}. Currently, research on rerouting attacks is limited, with only two related works~\cite{shafran2025rerouting,lin2025life} covering three attack methods. 
Take the cost escalation scenario as an example, Shafran et al.~\cite{shafran2025rerouting} utilize the win rate calculated by LLM router~(i.e., under gray-box access) to iteratively optimize the trigger to maximize $P_{\theta}(\text{win}{\text{s}} \mid t)$, where $t$ represents the crafted trigger. 
In the white-box setting described in~\cite{lin2025life}, the attacker first constructs a training set and uses the gradient information from the router to maximize $P_{\theta}(\text{win}_{\text{s}} \mid t \oplus q)$, where $q$ represents normal queries from the training set. Specifically, the dataset is composed of the Chatbot Arena dataset~\cite{zheng2023judging} and 30\% of the GPT-4 judge dataset~\cite{ong2024routellm}. Besides, due to requirements for gradient inofrmation, the white-box attack can only directly apply to limited routers~(such as $R_{LLM}$), but the optimized triggers can transfer to other routers.
In the box-free setting of \cite{lin2025life}, the attacker uses a set of proxy routers to select queries that have high strong model win rates across all routers. After that, they use a summarizer model to extract features and craft triggers. Following~\cite{lin2025life}, we set the proxy routers as the four evaluated routers and the summarizer model as GPT-4o.
For each attack, we follow the official settings and repeat the experiment 10 times~(i.e., 10 triggers per method) to evaluate the effectiveness of each attack method.



\textbf{Metrics}. We define the following metrics:

\begin{itemize}
	\item \textbf{Benchmark Score}: The performance of the routing system on different benchmarks. For MMLU and GSM8K, where each question has a standard answer, given a set of queries $Q = \{q_1, q_2, \dots, q_N\}$, the benchmark score is calculated as:
	\begin{equation}
		\text{Benchmark Score} = \frac{\sum_{i=1}^{N} \mathbb I(r_i = \hat{r}_i)}{N} \times 100
	\end{equation}
	
	where $r_i$ and $\hat{r}_i$ are the standard and predicted answers of routed model for query $q_i$, respectively. For MT-Bench, which only has reference answers, we use GPT-4o to evaluate response quality~(score 1-10). 
	
	\item \textbf{Rerouting Attack Success Rate (ASR)}: The effectiveness of an adversarial trigger in changing the model routing behavior. For example, in the cost escalation scenario, ASR is the proportion of queries rerouted from $M_w$ to $M_s$ after adding the trigger:
	\begin{equation}
		\text{ASR} = \frac{\sum_{i=1}^{N} \mathbb I(P(M_s | q_i) < \alpha) \cdot \mathbb I(P(M_s | t \oplus q_i) \geq \alpha)}{\sum_{i=1}^{N} \mathbb I(P(M_s | q_i) < \alpha)}
	\end{equation}
	In quality hijacking and safety bypass scenarios, the definition is reversed.
	
	\item \textbf{Target Model Selection Rate}: The proportion of queries that are routed to the target model. In cost escalation scenario, the target model is $M_s$, while in quality hijacking and safety bypass scenarios, it is $M_w$:
	\begin{equation}
		\text{Target Model Selection Rate} = \frac{\sum_{i=1}^{N} \mathbb I(P(M_t | q_i) \geq \alpha)}{N}
	\end{equation}
	
	For normal queries, the selection rate is 0.5 because we set $\alpha$ as the median of the overall win rate distribution of the dataset.
	
	\item \textbf{Adversarial Confidence Gain (ACG)}~\cite{lin2025life}: The change in strong model win rates after applying the adversarial attack:
	\begin{equation}
		\text{ACG} = \frac{1}{N} \sum_{i=1}^{N} \left( P(M_s | t \oplus q_i) - P(M_s | q_i) \right)
	\end{equation}
	
	\item \textbf{Jailbreak Success Rate~(JSR)}: For Jailbreak dataset, we use the Llama-2-13B classifier from HarmBench~\cite{mazeika2024harmbench} to evaluate whether the response of the routed LLM is harmful. Formally, JSR is defined as: 
	\begin{equation}
		\text{JSR} = \frac{\sum_{i=1}^{N} \mathbb I(J(r_i)=1)}{N}
	\end{equation}
	where $J$ is the jailbreak judge.
\end{itemize}

\subsection{Experimental Results}
\label{sec:rq2-results}

\subsubsection{Cost Escalation}
\label{sec:rq2-threat1}

To investigate whether existing LLM routers are vulnerable to the cost escalation threat, we evaluate their robustness against three rerouting attack methods across three different benchmarks, i.e., MMLU, GSM8K, and MT-Bench. We present the ASR results for each router under different benchmarks in Figure~\ref{fig:threat1_benchmark_results}.

\begin{table}[ht!]
	\centering
	\caption{Summary of cost escalation attack results on different routers in GSM8K benchmark, where "Strong" means the strong model selection rate.}
	\label{table:threat1_gsm8k}
	\begin{tabular}{cccccc}
		\hline
		\textbf{Router} & \textbf{Attack} & \textbf{Score} & \textbf{ASR} & \textbf{Strong} & \textbf{ACG} \\
		\hline
		\multirow{3}{*}{$R_{CLS}$} & No Attack & 34.0 &  & 50.0\% &  \\
		& Gray-box & 57.6 & 98.60\% & 99.3\% & 27.75\% \\
		& Box-free & 55.3 & 90.40\% & 94.8\% & 17.45\% \\
		\hline
		\multirow{3}{*}{$R_{MF}$} & No Attack & 38.0 &  & 50.0\% &  \\
		& Gray-box & 57.6 & 99.20\% & 99.6\% & 86.84\% \\
		& Box-free & 48.9 & 62.80\% & 79.8\% & 17.83\% \\
		\hline
		\multirow{3}{*}{$R_{SW}$} & No Attack & 38.0 &  & 50.0\% &  \\
		& Gray-box & 56.0 & 89.60\% & 94.8\% & 1.60\% \\
		& Box-free & 53.3 & 79.40\% & 89.7\% & 1.08\% \\
		\hline
		\multirow{4}{*}{$R_{LLM}$} & No Attack & 33.0 &  & 50.0\% &  \\
		& Gray-box & 54.9 & 85.80\% & 92.9\% & 32.72\% \\
		& Box-free & 44.4 & 41.00\% & 70.2\% & 9.58\% \\
		& White-box & 50.4 & 69.20\% & 84.6\% & 20.70\% \\
		\hline
	\end{tabular}
\end{table}

\textbf{Overall Performance.} The experimental results show that existing LLM routers are highly susceptible to the cost escalation threat, especially $R_{CLS}$ and $R_{SW}$, which have an ASR close to 90\% in gray-box attack across all benchmarks. In contrast, the $R_{MF}$ is relatively safer, with an ASR below 70\% in MMLU benchmark. However, it also showed high vulnerability on the other two benchmarks, even though the used triggers in all benchmarks are totally same~(i.e., universal triggers), which indicates that the router's vulnerability is also dependent on query types. Specifically, MT-Bench, shows the highest ASR across all routers, while GSM8K~(math problems) and MMLU~(multiple-choice questions) demonstrate relatively lower ASRs. This indicates that open-ended questions may be more susceptible to adversarial manipulation compared to more complex query types.

\begin{table}[ht!]
	\centering
	\caption{Summary of quality hijacking attack results on different routers in GSM8K benchmark, where "Weak" means the weak model selection rate.}
	\label{table:threat2_gsm8k}
	\begin{tabular}{cccccc}
		\hline
		\textbf{Router} & \textbf{Attack} & \textbf{Score} & \textbf{ASR} & \textbf{Weak} & \textbf{ACG} \\
		\hline
		\multirow{3}{*}{$R_{CLS}$} & No Attack & 34.0 &  & 50.0\% &  \\
		& Gray-box & 18.1 & 74.20\% & 86.9\% & -10.27\% \\
		& Box-free & 25.9 & 38.80\% & 67.1\% & -3.50\% \\
		\hline
		\multirow{3}{*}{$R_{MF}$} & No Attack & 38.0 &  & 50.0\% &  \\
		& Gray-box & 15.1 & 94.80\% & 97.4\% & -37.17\% \\
		& Box-free & 18.7 & 77.40\% & 88.7\% & -22.10\% \\
		\hline
		\multirow{3}{*}{$R_{SW}$} & No Attack & 38.0 &  & 50.0\% &  \\
		& Gray-box & 21.5 & 71.40\% & 85.4\% & -0.95\% \\
		& Box-free & 35.0 & 17.80\% & 56.6\% & -0.20\% \\
		\hline
		\multirow{4}{*}{$R_{LLM}$} & No Attack & 33.0 &  & 50.0\% &  \\
		& Gray-box & 33.8 & 10.40\% & 50.9\% & 0.50\% \\
		& Box-free & 25.7 & 29.80\% & 64.9\% & -6.90\% \\
		& White-box & 34.3 & 8.60\% & 47.8\% & 1.36\% \\
		\hline
	\end{tabular}
	\vspace{-5pt}
\end{table}

\textbf{Performance across Different Attack Methods}. Among these attack methods, the gray-box attack shows the best performance, with an ASR of more than 85\% in most cases. Box-free attack, on the other hand, is more average and shows distinct performance across different routers. For example, it has an ASR of more than 75\% in $R_{CLS}$ and $R_{SW}$, but shows extremely low ASR in $R_{MF}$, ranging from 23.4\% to 62.8\%. White-box attack, specific to $M_{LLM}$ due to gradient constrict, shows similar performance to gray-box attack on MMLU and MT-Bench, but relatively low in GSM8K. The reason is that, unlike gray-box attack that directly optimizes triggers, white-box attack optimizes the triggers on a training dataset~(real-world open-ended questions). Therefore, they offer excellent transferability on similar datasets like MMLU and MT-Bench, but relatively limited transferability on mathematical problems like GSM8k.

\begin{figure*}[t]
	\centering
	\begin{subfigure}{0.32\textwidth}
		\centering
		\includegraphics[width=\textwidth]{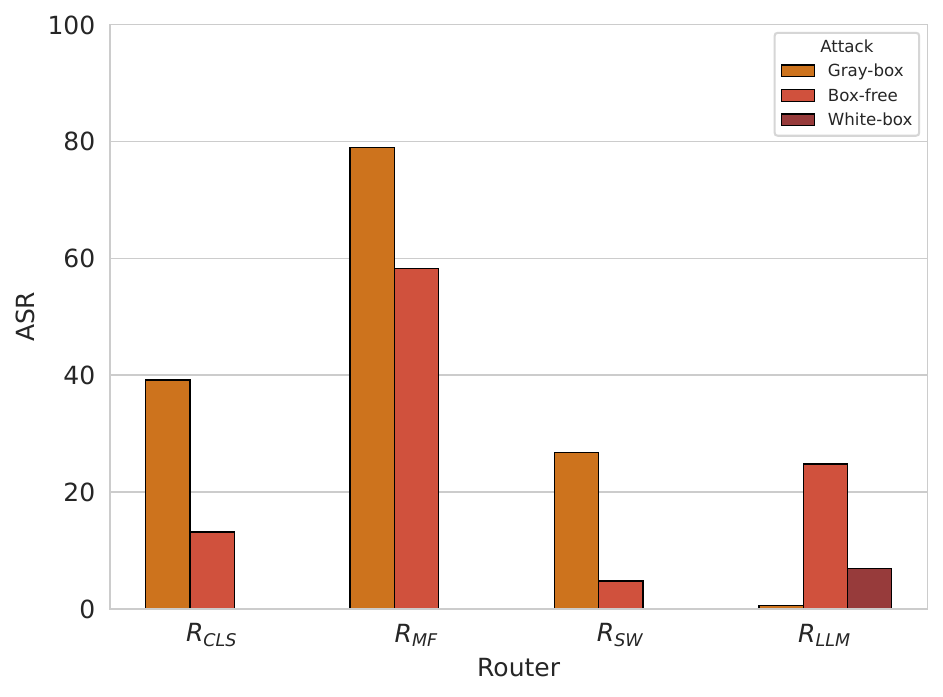}
		\subcaption{MMLU} \label{fig:threat2_mmlu}
	\end{subfigure}%
	\hspace{0.01\textwidth}
	\begin{subfigure}{0.32\textwidth}
		\centering
		\includegraphics[width=\textwidth]{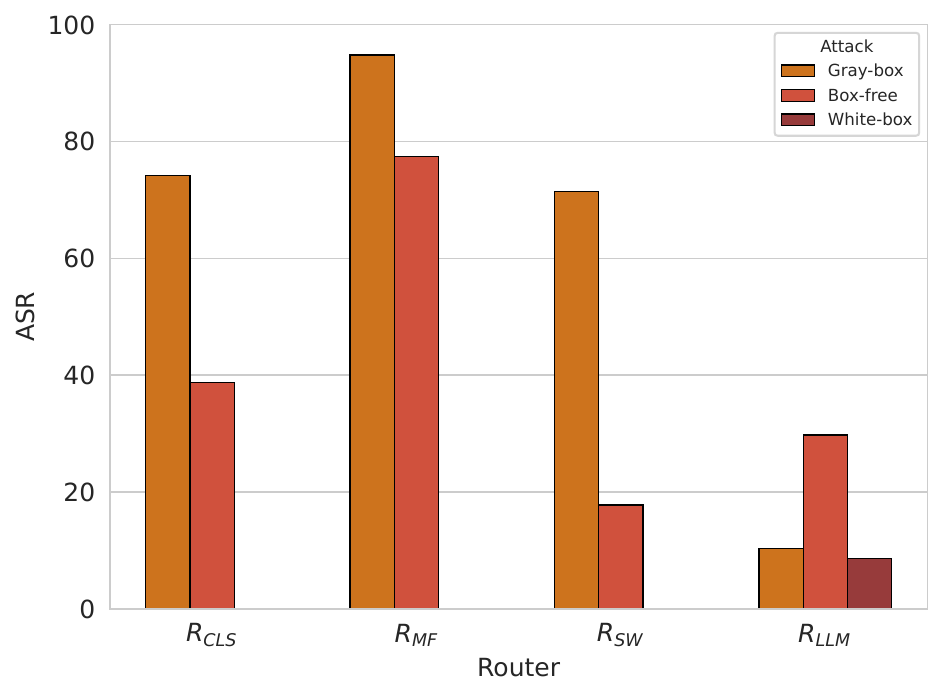}
		\subcaption{GSM8K} \label{fig:threat2_gsm8k}
	\end{subfigure}%
	\hspace{0.01\textwidth}
	\begin{subfigure}{0.32\textwidth}
		\centering
		\includegraphics[width=\textwidth]{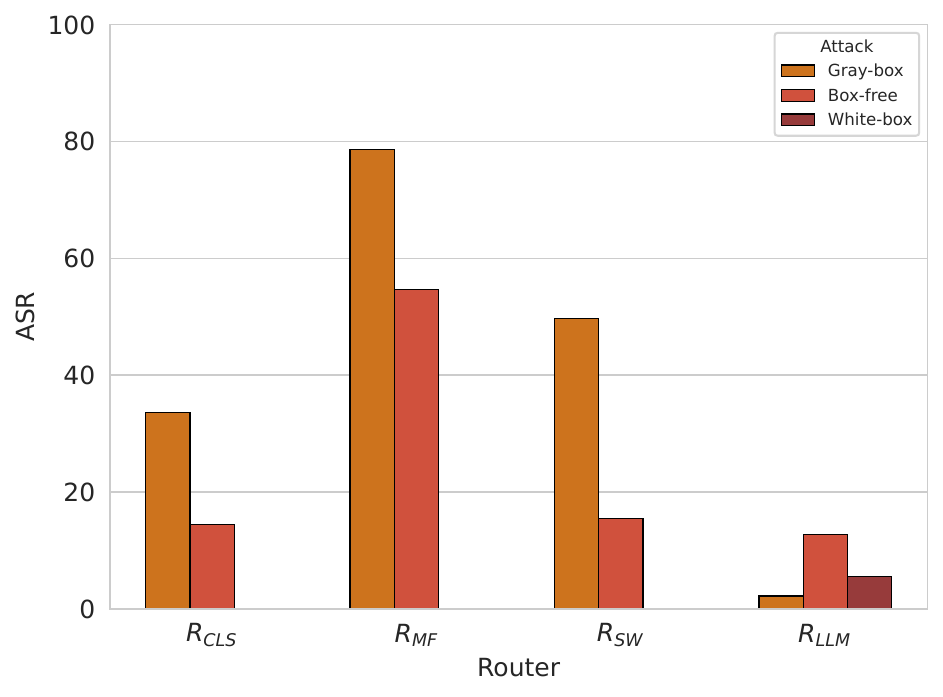}
		\subcaption{MT-Bench} \label{fig:threat2_mt_bench}
	\end{subfigure}
	\caption{The quality hijacking ASR of different attack methods in four routers in three benchmarks. Similar to cost escalation, the $M_s$ is GPT-4o, while the $M_w$ is Mixtral-8x7B-Instruct.} 
	\label{fig:threat2_benchmark_results}
\end{figure*}

\textbf{Response Quality and Cost Analysis}. To explore the changes in cost and response quality after rerouting attacks, we present the complete results of various metrics on GSM8K in Table~\ref{table:threat1_gsm8k}, where the performance of $M_s$ is almost four times that of $M_w$~(58 vs. 14 in Table~\ref{table:threshold_and_performance}). Experimental results show that rerouting attacks improve response quality, i.e., the higher the ASR, the more queries will be routed to the strong model, and thus the higher benchmark score. For example, when the ASR is 99.2\%, the benchmark score is 57.6, which is almost the same as the performance of $M_s$. However, the required cost is also greatly increased. Note that the $M_s$ we use is the GPT-4o model (requires API overhead), while $M_w$ is a locally deployed Mistral-8x7B-Instruct~(no API cost). Therefore, ASR directly reflects the increase in cost. When the ASR is 100\%, the required cost is directly doubled. In addition, the API cost will increase rapidly with the increase in the number of questions, and the attacker only needs to construct a universal trigger, which highlights the threat of this attack.

\subsubsection{Quality Hijacking}
\label{sec:rq2-threat2}

To explore whether LLM routers exhibit similar vulnerabilities to the quality hijacking threat, we change the attack goal to reroute to $M_w$ to downgrade task performance. The used routers, benchmarks, and rerouting attack methods are the same as in Section~\ref{sec:rq2-threat1}. The ASR results of each attack under different benchmarks are shown in Figure~\ref{fig:threat2_benchmark_results}.

\textbf{Overall Performance}. Overall, the threat of quality hijacking is less significant than that of cost escalation. No single ASR reaches 100\%, with the ASR remaining below 60\% in most cases. Interestingly, the $R_{MF}$ that is relatively safe in cost escalation, exhibited the greatest vulnerability to quality hijacking (with ASRs exceeding 54\% across all attack methods on all benchmarks). 

The threat of quality hijacking also depends on the question type. Specifically, the GSM8K (high school math problems) is the most susceptible to triggers. For instance, on GSM8K, $R_{MF}$'s ASR for both attack methods exceeds 77\%, the highest across all scenarios. It is worth noting that, as shown in Table 2, the 50th percentile threshold for gsm8k is the highest across all three benchmarks for nearly all routers, i.e., the overall $P_{\theta}(\text{win}_{\text{s}}$ of gsm8k (which can represent the difficulty of the problem) is higher than the other two. Therefore, we speculate that difficult queries may be more susceptible to quality hijacking, which further highlights the severity of this threat.

\textbf{Performance across Different Attack Methods}. The results show that the gray-box attack outperforms other attack methods in all routers, except for $R_{LLM}$. Specifically, in MMLU and MT-Bench, the ASR of the gray-box attack is close to 0, and the selection rate of $M_w$ even decreases from 50\% to 32.5\% and 42.9\%, respectively. In other words, after applying these triggers, user queries are more likely to be routed to strong models, which is entirely contrary to the intended goal of the attack. We attribute this phenomenon to the fact that adding triggers inherently increases the complexity of the query, particularly with the introduction of rare words or higher perplexity. For example, in GSM8K, only the gray-box and white-box attacks (with garbled triggers) sometimes increase the strong model win rate (ACG > 0), as shown in Table~\ref{table:threat2_gsm8k}, while the box-free attack (with semantic triggers) consistently reduces the strong model win rate, although the effect is relatively small.

\textbf{Response Quality and Cost Analysis}. We present the detailed attack results for different routers and attack methods at the GSM8K benchmark in Table~\ref{table:threat2_gsm8k}, where the performance of $M_s$ is four times that of $M_w$.. The results demonstrate that the quality hijacking attack reduces the response quality of the LLM router system to some extent. For instance, the gray-box and box-free attacks reduce the benchmark score by 60.3\% and 50.8\%, respectively, in $R_{MF}$. Given the universality of these triggers and the fact that complex queries~(e.g., math problems) are particularly vulnerable to such attacks, this threat should not be underestimated.

\begin{table}[t]
	\centering
	\caption{Summary of safety bypass attack results on different routers in Jailbreak benchmark.}
	\label{table:threat3_jailbreak}
	\begin{tabular}{cccccc}
		\hline
		\textbf{Router} & \textbf{Attack} & \textbf{JSR} & \textbf{ASR} & \textbf{Weak} & \textbf{ACG} \\
		\hline
		\multirow{3}{*}{$R_{CLS}$} & No Attack & 45.2 &  & 50.0\% &  \\
		& Gray-box & 52.4 & 23.36\% & 61.4\% & -9.04\% \\
		& Box-free & 47.3 & 13.52\% & 54.0\% & -3.79\% \\
		\hline
		\multirow{3}{*}{$R_{MF}$} & No Attack & 42.8 &  & 50.0\% &  \\
		& Gray-box & 73.4 & 92.40\% & 96.2\% & -47.10\% \\
		& Box-free & 63.7 & 62.56\% & 81.3\% & -27.30\% \\
		\hline
		\multirow{3}{*}{$R_{SW}$} & No Attack & 45.2 &  & 50.4\% &  \\
		& Gray-box & 58.9 & 44.84\% & 72.6\% & -1.39\% \\
		& Box-free & 46.7 & 10.16\% & 53.3\% & -0.24\% \\
		\hline
		\multirow{4}{*}{$R_{LLM}$} & No Attack & 40.4 &  & 50.0\% &  \\
		& Gray-box & 39.5 & 8.72\% & 47.8\% & 2.69\% \\
		& Box-free & 50.8 & 35.84\% & 67.0\% & -11.07\% \\
		& White-box & 39.0 & 5.52\% & 46.6\% & 3.35\% \\
		\hline
	\end{tabular}
	\vspace{-5pt}
\end{table}

\subsubsection{Safety Bypass}

To investigate whether existing LLM routers are susceptible to safety bypass threat, we further conduct experiments on Jailbreak benchmark. The used routers and routing attack methods are the same as Section~\ref{sec:rq2-threat1} and Section~\ref{sec:rq2-threat2}. We present the detailed attack results in Table~\ref{table:threat3_jailbreak}.

\textbf{Overall Performance}. The experimental results are very similar to those of the quality hijacking threat, especially the MT-Bench benchmark, which also consists of open-ended questions. However, the ASR on the Jailbreak benchmark is relatively higher. Specifically, the $R_{MF}$ exhibits the highest vulnerability, with an ASR above 60\% for all attack methods, whereas the ASR for other routers is below 45\%. 

Among the different attack methods, the gray-box attack still demonstrates relatively better performance, except for the $R_{LLM}$, where the box-free attack achieves a high ASR (35.84\%), while the white-box attack shows almost no effectiveness (ASR of 5.52\%).

\begin{figure}[t]
	\centering
	\includegraphics[width=.5\textwidth]{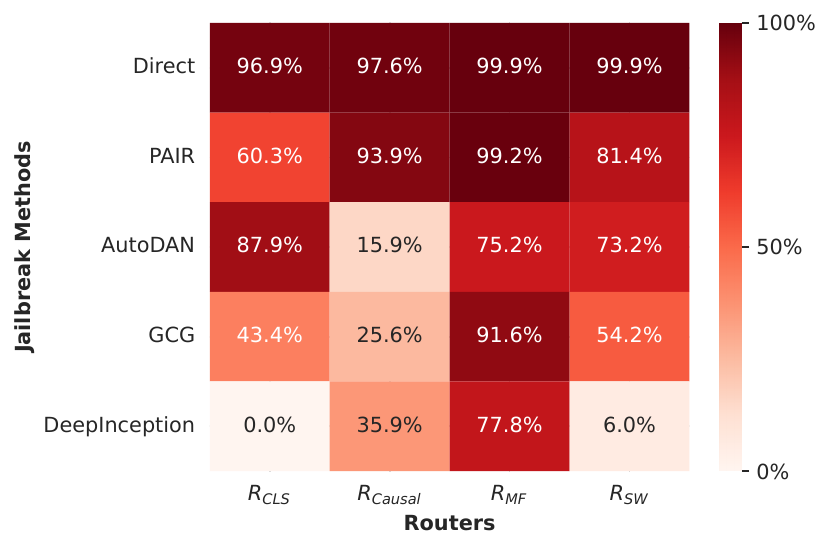}
	\caption{The heatmap of safety bypass ASR across 5 types of jailbreak prompts against 4 LLM routers. We average the ASR in three rerouting attack methods.}
	\vspace{-10pt}
	\label{fig:jailbreak_heatmap}
\end{figure}

\textbf{Performance across Different Jailbreak Attack Methods}. To explore whether different types of jailbreak prompts affect the effectiveness of rerouting attacks, we plot a heatmap of the rerouting ASR values of different jailbreak prompts in different routers, as shown in Figure~\ref{fig:jailbreak_heatmap}. The experimental results demonstrate that short and semantically meaningful jailbreak methods, such as Direct and PAIR are more likely to be affected by rerouting methods. 
On the other hand, AutoDAN and GCG are relatively less susceptible to rerouting attacks, possibly because the lengthy jailbreak template and unreadable jailbreak trigger suffix affect the effectiveness of the rerouting trigger.
Notably, the jailbreak prompts of DeepInception, which uses nested hypnotic scenarios to induce the LLM to jailbreak, are least susceptible to rerouting attacks, especially in $R_{CLS}$ and $R_{SW}$~(ASR close to 0). 

\begin{figure}[t]
	\centering
	\includegraphics[width=.4\textwidth]{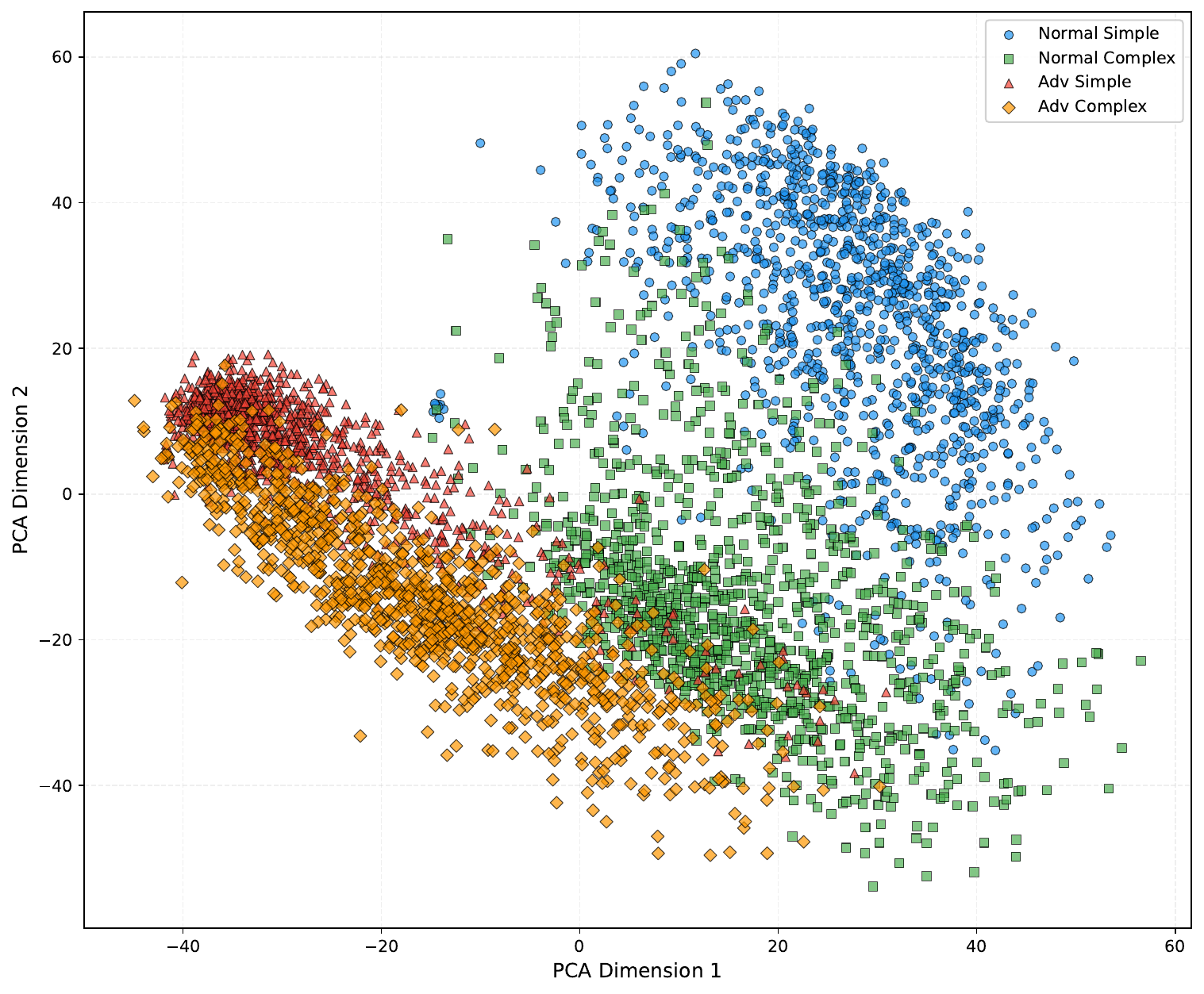}
	\caption{PCA visualization of query representations.}
	\label{fig:visualization_causal_llm}
\end{figure}

\textbf{Response Safety Analysis}. Similar to quality hijacking threat, there are more harmful queries rerouted to the less safe $M_w$ after applying adversarial attacks, resulting in higher JSR. For example, the gray-box attack has an ASR of 92.4\% in $R_{MF}$, which increases the JSR from 42.8\% to 73.4\%. This highlights the security threat rerouting attacks pose to multi-model systems (such as GPT-5) with LLM routers, as malicious user could hijack the router to route to weaker and less secure models to obtain jailbreak responses. Furthermore, considering the "water bucket" effect, a system's security ultimately depends on its weakest model. If rerouting attack is combined with other attacks such as worm attacks, i.e., using weak models as a springboard to compromise other models, it could cause incalculable damage.


\subsection{Attack Analysis}
\textbf{Mechanism Analysis}. To investigate the underlying reasons of the effectiveness of rerouting attacks, particularly in the cost escalation threat, we analyze their semantic representations. Specifically, we consider the custom dataset from \cite{lin2025life} that includes 38547 real-world open-ended questions. We calculate the strong model win rates $P_{\theta}(M_s \mid q)$ for each query and then sample two kinds of queries: 1129 "simple" queries whose $P_{\theta}(M_s \mid q)$ consistently fall below the router threshold, and 1156 "complex" queries whose $P_{\theta}(M_s \mid q)$ are consistently above the router threshold. We prepend rerouting triggers to these queries, resulting in four data groups: normal simple, normal complex, adversarial simple, and adversarial complex. 
We then extract the last-layer hidden states of $R_{LLM}$ for these queries, which represent the internal features used by the router to make decisions. We apply PCA to visualize the distribution of these representations. As shown in Figure~\ref{fig:visualization_causal_llm}, rerouting prompts form a compact cluster in the embedding space, clearly separated from benign simple queries but closer to benign complex queries. This suggests that the router perceives rerouting queries as having a similar level of complexity to normal complex queries, which explains why rerouting attacks are successful.

\begin{figure}[t]
	\centering
	\includegraphics[width=.45\textwidth]{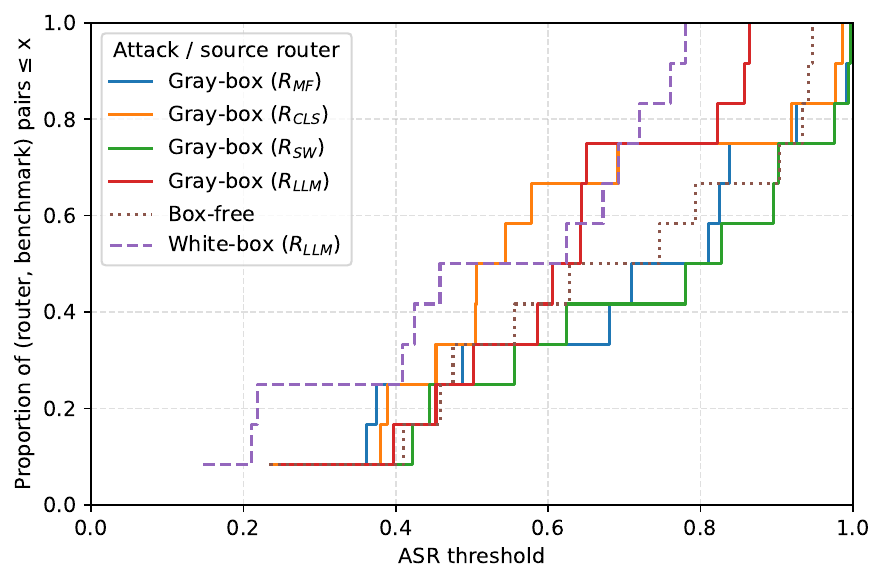}
	\caption{CDF of ASR values for six groups of triggers across different benchmarks and routers.}
	\label{fig:cdf_all_attacks_strong}
\end{figure}

\begin{figure}[t]
	\centering
	\includegraphics[width=.45\textwidth]{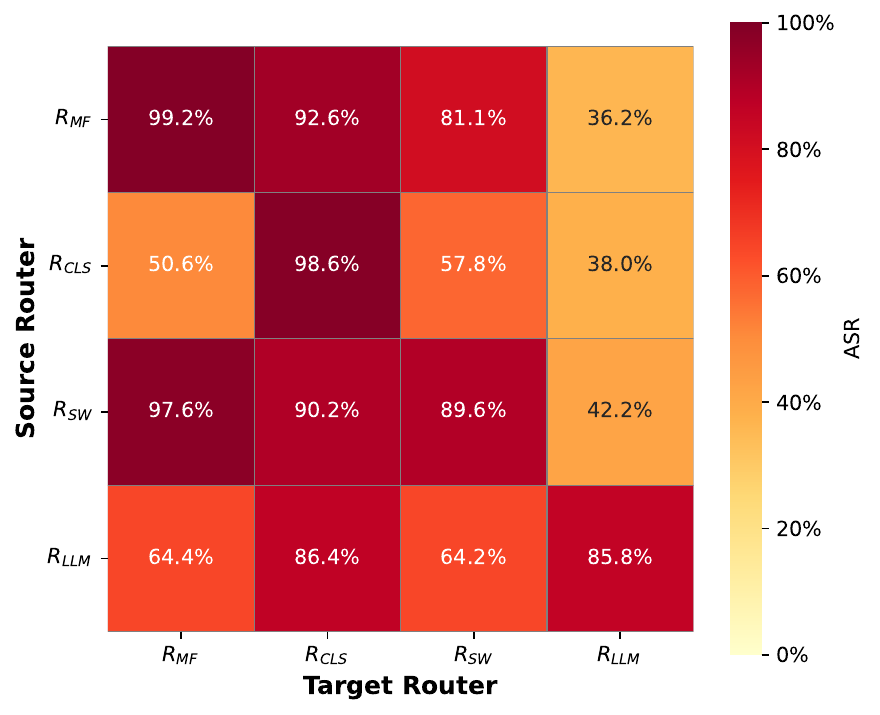}
	\caption{Transfer results of gray-box attack on GSM8K benchmark across different routers.}
	\label{fig:transfer_gsm8k_strong}
\end{figure}

\begin{figure*}[ht!]
	\centering
	\includegraphics[width=.94\textwidth]{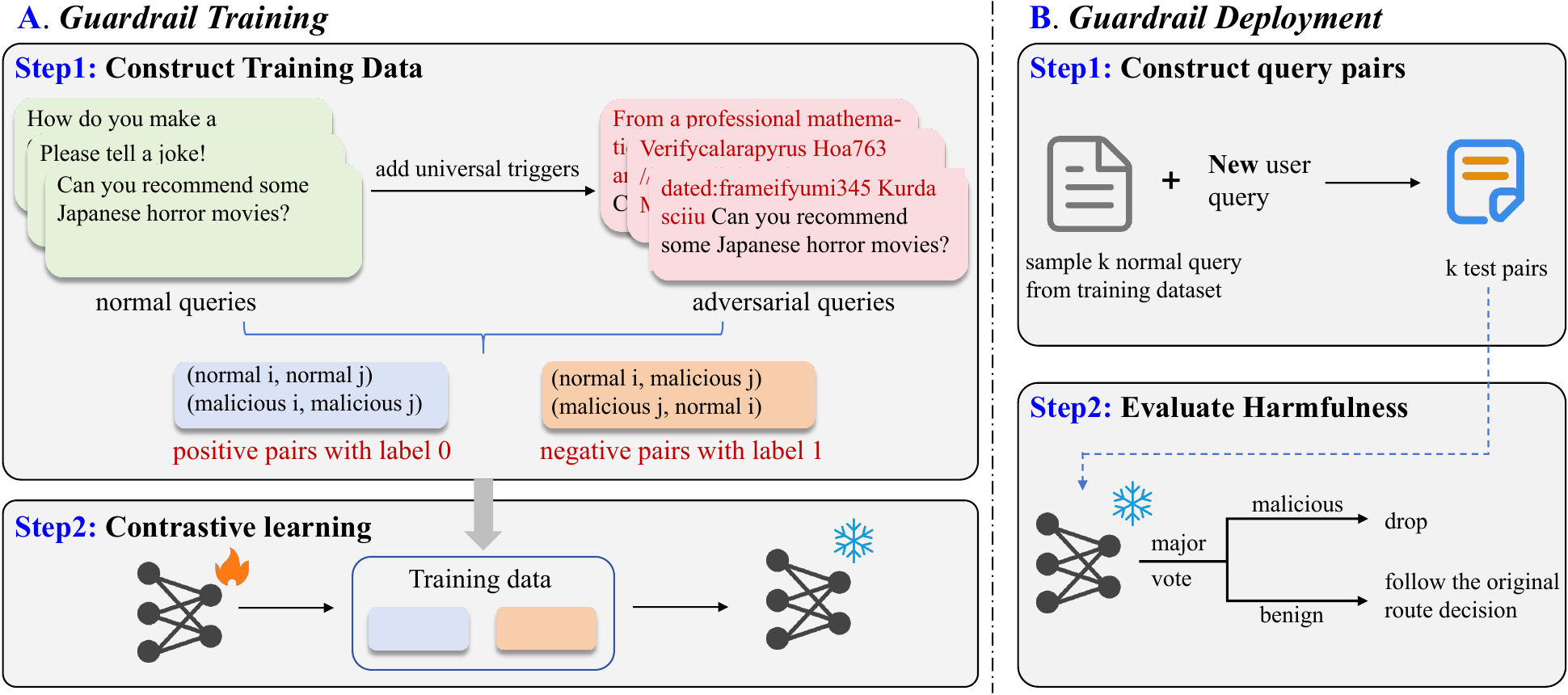}
	\caption{An overview of RerouteGuard, which is a plug-and-play rerouting detector that could detect and filter suspicious prompts before passing it to LLM routers. Specifically, during training, it utilize contrastive learning to learn distinctive semantic patterns that characterize adversarial rerouting prompts. In the deployment phase, the trained model operates as a real-time filter and blocks adversarial rerouting prompts before they can interfere with the LLM routing decision process.}
	\label{fig:framework}
\end{figure*}


\textbf{Transferability}. We analyze the transferability of the rerouting triggers across different benchmarks and routers. Specifically, we apply the six groups of triggers, as described in Section~\ref{sec:rq2-threat1}, to three benchmarks and four routers, resulting in 12 ASR values for each group. We then analyze the distribution of these ASR values by calculating the Cumulative Distribution Function (CDF), as shown in Figure~\ref{fig:cdf_all_attacks_strong}. The results indicate that most triggers exhibit moderate transferability, with over half of the 12 cross-benchmark and cross-router settings achieving an ASR greater than 0.6, though the transfer performance remains inconsistent across targets.

We further investigate the transferability of triggers across different routers. We present the cross-router ASR results of the gray-box attack in GSM8K benchmark in Figure~\ref{fig:transfer_gsm8k_strong}. The results show that triggers optimized on $R_{MF}$ and $R_{SW}$ exhibit good transferability to other routers except for $R_{LLM}$. Notably, $R_{CLS}$ is the most susceptible router, with all transfer ASR values exceeding 86\%.

\textbf{Trigger Patterns}.
To investigate whether there is a significant syntactic difference between benign prompts and rerouting prompts, we calculate the perplexity and length (measured by number of characters) of each prompt. The experimental results are presented in Table~\ref{table:trigger_pattern}. It can be observed that rerouting triggers, particularly those that are "unreadable,"~(e.g., gray-box and white-box attack) result in a significant increase in perplexity, while semantically valid triggers do not lead to such an increase, as they maintain a reasonable level of coherence with the original prompt. 

Regarding prompt length, rerouting prompts are slightly longer than normal prompts, as triggers are appended directly before the original query. This increase is minimal in the case of the gray-box and white-box attack scenarios, where length constraints are typically imposed during optimization. In contrast, box-free triggers tend to be longer, due to fewer restrictions on their generation~(i.e., generated by LLM). This indicates that rerouting prompts are difficult to detect and mitigate using simple rule-based defense mechanisms, as they can be easily bypassed by imposing optimization constraints.

\begin{table}[t]
	\centering
	\caption{The perplexity and prompt length of normal prompts and rerouting prompts.}
	\label{table:trigger_pattern}
	\begin{tabular}{cccc}
		\hline
		\textbf{Benchmark} & \textbf{Setting} & \textbf{PPL} & \textbf{Prompt Length} \\
		\hline
		\multirow{4}{*}{MMLU} & Normal & 26.5 & 482.3 \\
		& Gray-box attack & 74.4 & 528.8 \\
		& Box-free attack & 32.5 & 731.6 \\
		& White-box attack & 64.4 & 527.9 \\
		\hline
		\multirow{4}{*}{GSM8K} & Normal & 38.6 & 333.6 \\
		& Gray-box attack & 113.0 & 380.1 \\
		& Box-free attack & 44.4 & 582.9 \\
		& White-box attack & 107.2 & 379.2 \\
		\hline
		\multirow{4}{*}{MT-Bench} & Normal & 49.2 & 318.3 \\
		& Gray-box attack & 330.8 & 364.8 \\
		& Box-free attack & 51.7 & 567.6 \\
		& White-box attack & 334.0 & 363.9 \\
		\hline
	\end{tabular}
\end{table}

\section{RerouteGuard}

To address \textbf{RQ3}, we propose \textit{RerouteGuard}, a novel plug-and-play defense mechanism that identifies and blocks malicious rerouting prompts before they reach the LLM router, thereby preserving the integrity of the routing decision process without modifying the router architecture. 
In Section~\ref{sec:overview}, we provide an overview of our defense mechanism. We then introduce the construction of the training dataset and the key designs of \textit{RerouteGuard} in Section~\ref{sec:data-preparation} and Section~\ref{sec:train}, respectively. In Section~\ref{sec:deployment}, we present how it operates on new queries. 

\subsection{Overview}
\label{sec:overview}
Building on the important observation from Figure~\ref{fig:visualization_causal_llm} that adversarial rerouting prompts exhibit distinctive patterns in their semantic representations, our method trains a siamese network to distinguish between normal queries and rerouting prompts through contrastive learning. As shown in Figure~\ref{fig:framework}, the overall pipeline of RerouteGuard comprises three main phases: data preparation, guardrail training, and guardrail deployment. In the data preparation phase, we construct a carefully balanced dataset of query pairs, including both positive pairs (normal-normal or adversarial-adversarial) and negative pairs (normal-adversarial). The adversarial examples are generated by prepending malicious triggers to normal queries, with a sophisticated sampling strategy that ensures coverage of diverse attack patterns. During model training, we employ a siamese network architecture trained with a composite loss function that combines binary classification and contrastive learning objectives. This enables the model to learn distinctive semantic patterns that characterize adversarial rerouting attempts. Finally, in the deployment phase, the trained model operates as a real-time filter, processing new user queries and blocking those identified as adversarial before they can interfere with the LLM routing decision process.

\subsection{Semantic-guided Data Preparation}
\label{sec:data-preparation}
For normal queries, we use the custom dataset from~\cite{lin2025life}, which includes 38,547 real-world open-ended questions drawn from the Chatbot Arena dataset~\cite{zheng2023judging} and the GPT-4 judge dataset~\cite{ong2024routellm}. We denote this dataset as $D_{raw}$. We then employ four distinct routers~(i.e., $R_{CLS}$, $R_{MF}$, $R_{SW}$, $R_{LLM}$) to compute the strong model win rate for each query in $D_{raw}$. Based on these win rate, we select queries whose win rate are consistently within the bottom 50\% across all routers. This yields 11,557 simple questions referred to as $D_{normal}$, which is then partitioned into training, validation, and test subsets with an 8:1:1 ratio. The reason is that, the cost escalation scenario aims to reroute simple queries to strong models, so we believe that using simple queries to train the guardrails will yield more effective results.

For adversarial queries, we generate 50 triggers for each rerouting method, referred to as $\mathcal{T}$, and split them into train/validation/test subsets with a 6:2:2 ratio.
To generate rerouting examples, for each query in a given data split of $D_{normal}$, we randomly sample one trigger from the corresponding trigger subset and prepend it to the user query, producing the malicious query used for training, validation, or testing. This per‑query random pairing ensures diversity in trigger‑query combinations and prevents data leakage. We refer to this dataset as $D_{adv}$.

Based on $D_{normal}$ and $D_{adv}$, we construct a carefully balanced dataset comprising positive and negative pairs. Specifically, positive pairs consist of either normal-normal query pairs or adversarial-adversarial pairs that share the same semantic distribution, thus teaching the model to recognize queries from identical categories. In contrast, negative pairs combine normal queries with adversarial variants. We implement a sophisticated sampling strategy that balances two types of negative pairs: cross-index negatives that pair a normal query with adversarial variants of different queries, and self-index negatives that pair a normal query with its own adversarial variants, with the latter serving as particularly challenging examples. This balanced approach, controlled by a \textit{negative cross ratio} hyperparameter, ensures the model learns to recognize both inter-query and intra-query adversarial patterns.

\subsection{Contrastive-learning-based Guardrail Training}
\label{sec:train}
The core of RerouteGuard employs a dual-encoder siamese architecture with shared parameters. Specifically, we utilize a pre-trained transformer model as its backbone encoder, which processes an input query pair $p_i = (q_a, q_b)$ through CLS token pooling. A projection head then maps the encoder outputs to a normalized embedding space, while a pair classifier combines the concatenated embeddings, their absolute differences, and element-wise products to produce final binary classifications:

\begin{equation}
	e_a = \text{Encoder}(q_a), \quad e_b = \text{Encoder}(q_b)
\end{equation}

\begin{equation}
	\text{logits} = \text{Classifier}(e_a, e_b)
\end{equation}


We design a multi-objective loss function that strategically combines binary classification with contrastive regularization:

\begin{equation}
	\label{eq:loss}
	\mathcal{L} = \lambda_{bce} \cdot \mathcal{L}_{BCE} + \lambda_{contr} \cdot \mathcal{L}_{SupCon}
\end{equation}

The binary cross-entropy loss function, $\mathcal{L}_{BCE}$, is used to measure the discrepancy between the predicted logits and the ground truth labels~(0 for positive pairs and 1 for negative pairs). It is defined as:

\begin{equation}
	\label{eq:bce}
	\mathcal{L}_{BCE} = - \frac{1}{N} \sum_{i=1}^{N} \left[ y_i \log(\hat{y_i}) + (1 - y_i) \log(1 - \hat{y_i}) \right]
\end{equation}

where $N$ is the number of samples in the batch, $y_i$ is the true label for pair $p_i$ (0 or 1), and $\hat{y_i}$ is the predicted probability for $p_i$.

The supervised contrastive loss, $\mathcal{L}_{SupCon}$, encourages the model to learn embeddings where samples from the same class are closer in the embedding space, while samples from different classes are farther apart:

\begin{equation}
	\label{eq:supcon}
	\mathcal{L}_{SupCon}=-\frac{1}{2N} \sum_{i=1}^{2N} \frac{\sum_{j \in \mathcal{P}_i} \cdot \log \left(\frac{\exp(\text{sim}(e_i, e_j) / \tau)}{\sum_{k \in \mathcal{D}_i} \exp(\text{sim}(e_i, e_k) / \tau) \cdot w_{i k}}\right)}{\sum_{j \in \mathcal{P}_i} 1}
\end{equation}

where $N$ is the batch size, $\text{sim}(e_i, e_j) = \frac{e_i^\top e_j}{\|e_i\| \|e_j\|}$ is the cosine similarity between embeddings $e_i$ and $e_j$, $\tau$ is the temperature hyperparameter that controls the scale of the similarities, $\mathcal{P}_i$ is the set of positive samples for anchor sample $i$:
\begin{equation}
	\mathcal{P}_i = \mathbb{I}(y_i = y_j), \quad \forall i, j \in \{1, 2, \dots, 2N\}
\end{equation}
$\mathcal{D}_i$ is the denominator mask that determines which pairs of embeddings (sample $i$ and sample $j$) are considered when computing the denominator of the loss function:
\begin{equation}
	\mathcal{D}_i = \mathcal{P}_i \cup \mathbb{I}\left(\text{valid\_pair}(i, j)\right), \quad \forall i, j \in \{1, 2, \dots, 2N\}
\end{equation}
Specifically, $w_{ij}$ is the weight applied to the pair $(i, j)$, where misclassified negative pairs receive higher weighting in the contrastive denominator. This approach ensures that the model develops robust discrimination capabilities between normal and adversarial examples, especially the self-index negative pairs where the rerouting queries are highly similar to the normal queries.

\begin{algorithm}
	\caption{RerouteGuard Defense Mechanism}
	\label{alg:rerouteguard}
	\begin{algorithmic}[1]
		\Procedure{DataPreparation}{$D_{\text{raw}}$}
		\State $D_{\text{normal}} \gets \{q \in D_{\text{raw}} \mid \text{win-rate}(q) \in \text{bottom }50\%\}$
		\State $D_{\text{normal}}^{\text{train}}, D_{\text{normal}}^{\text{val}}, D_{\text{normal}}^{\text{test}} \gets \text{split} (D_{\text{normal}})$
		\State $\mathcal{T}_{\text{train}}, \mathcal{T}_{\text{val}}, \mathcal{T}_{\text{test}} \gets \text{split} (\mathcal{T})$
		\For{$split \in \{\text{train}, \text{val}, \text{test}\}$}
		\State $D_{\text{adv}}^{split} \gets \text{construct\_adv}(D_{\text{normal}}^{split}, \mathcal{T}_{split})$
		\State $D_{split} \gets \text{construct\_pair}(D_{\text{normal}}^{split}, D_{\text{adv}}^{split})$
		\EndFor
		\State \Return $D_{\text{train}}, D_{\text{val}}, D_{\text{test}}$
		\EndProcedure
		
		\Procedure{TrainingPhase}{$D_{\text{train}}$}
		\State Initialize siamese network with pre-trained encoder
		\For{epoch $\gets 1$ to $E_{\text{max}}$}
		\For{batch $(q_a, q_b, y)$ in $D_{\text{train}}$}
		\State $e_a, e_b \gets \text{Encoder}(q_a), \text{Encoder}(q_b)$
		\State $\hat{y} \gets \text{Classifier}(e_a, e_b)$ 
		\State $\mathcal{L} \gets \lambda_{\text{bce}}\mathcal{L}_{\text{BCE}} + \lambda_{\text{contr}}\mathcal{L}_{\text{SupCon}}$ \Comment{Eq.~\eqref{eq:loss}}
		\State Update parameters with gradient clipping
		\EndFor
		\EndFor
		\State \Return Trained model $G$
		\EndProcedure
		
		\Procedure{DeploymentPhase}{$q_{\text{new}}, G, D_{\text{normal}}, K$}
		\State Sample $\{q_1, \dots, q_K\}$ from $D_{\text{normal}}^{\text{train}}$
		\State $\text{adv\_votes} \gets \sum_{i=1}^K \mathbb{I}(G(q_i, q_{\text{new}}) > 0.5)$ \Comment{0: normal, 1: adversarial}
		\State \textbf{return} $\text{adv\_votes} > K/2$ ? \textsc{Block} : \textsc{Forward}
		\EndProcedure
		
	\end{algorithmic}
\end{algorithm}

We also incorporate several advanced techniques to optimize performance during training. We employ discriminative learning rates that apply lower rates to backbone layers and higher rates to projection and classification heads, thus enabling stable feature extraction while allowing rapid adaptation in the decision layers. To prevent instability in the early stages of training, we adopt a contrastive warmup strategy, gradually increasing the contrastive loss weight. Additionally, we incorporate gradient clipping to mitigate gradient explosion and layer-wise decay to apply distinct weight regularization to the backbone and head parameters, ensuring balanced learning across the model.

\subsection{History-guided Guardrail Deployment}
\label{sec:deployment}

During deployment, RerouteGuard functions as an efficient preprocessing filter that operates in three sequential stages. First, given a new query $q$, RerouteGuard pairs it with $K$ reference queries from the normal query dataset $D_{normal}$, which results in $K$ query pairs:
\begin{equation}
	Q_{\text{pairs}} = \{ (q, q_i) \mid q_i \in D_{\text{normal}}, i = 1, 2, \dots, K \}
\end{equation}
The siamese network then computes classification probabilities for each of these query pairs. If a pair $(q, q_i)$ is classified as a positive pair, then $q$ is a normal query like $q_i$. Conversely, if the pair is classified as negative, then $q$ is determined to be adversarial. Finally, we perform a majority vote to determine the query label $\hat{y}$:
\begin{equation}
	\hat{y} = \arg\max_{c \in \{c_0, c_1\}} \sum_{i=1}^{K} \mathbb{I} (G((q, q_i)) = c)
\end{equation}
where $G$ is RerouteGuard, $c_0$ represents positive~(i.e., q is normal) and $c_1$ represents negative. Queries with high adversarial probability scores will be filtered out before reaching the LLM router. This plug-and-play characteristic of RerouteGuard allows seamless integration with existing LLM routing systems, providing an additional security layer without necessitating retraining of LLM routers.

\section{Experiments}
To evaluate the effectiveness of RerouteGuard, we conduct evaluations on four mainstream LLM routers against three existing rerouting attack methods. We first describe the experimental setup in Section~\ref{sec:defense-setting}, including the guardrail details, LLM routers, and attack methods. 
In Section~\ref{sec:defense-results}, we report and analyze the evaluation results to demonstrate the effectiveness of RerouteGuard. Finally, we present the ablation results of RerouteGuard in Section~\ref{sec:defense-ablation}.

\subsection{Experimental Settings}
\label{sec:defense-setting}
\textbf{Guardrail Details}. 
For data sampling, we generate 1 positive pair and 1 negative pair per normal sample. This results in a balanced mix of four pair types: benign-benign $(q_i, q_j)$, malicious-malicious $(m_i, m_j)$, and cross-domain pairs $(q_i, m_j)$ and $(m_j, q_i)$, with the latter generated by swapping the order of the cross-domain pair. We set the \textit{negative cross ratio} as 0.5, i.e., for each benign query $q_i$, there is a 50\% chance that $m_j$ will be selected as the corresponding rerouting query for $q_i$.
For guardrail training, we implement RerouteGuard using \texttt{XLM-RoBERTa-base}~\cite{facebookai2019xlm-roberta-base} as the text encoder, which is lightweight~(125MB) and has proven to be effective in multilingual text understanding, making it suitable for recognizing the complex patterns in adversarial triggers. We set the batch size to 64 and train the model using the AdamW optimizer with a learning rate of 3e-5. The maximum sequence length for tokenization is set to 256 tokens, and the pooling strategy for the output embeddings is set to \texttt{cls}, which uses the embedding of the \texttt{[CLS]} token as a sentence-level representation. Additionally, we set the weight of hard negative samples to 1.2 to emphasize the importance of challenging negative examples during training. The $\lambda_{bce}$ and $\lambda_{contr}$ are set to 0.65 and 0.35, respectively.
For guardrail deployment, we set the reference number $K$ to 4, i.e., when given a new query, RerouteGuard dynamically selects 4 queries from the training dataset to form pairs and make a decision through majority vote.

\textbf{Attack Methods and Datasets}.
The attack methods used are the same as in Section~\ref{sec:measurement}, i.e., white-box, gray-box and box-free settings.
Our test datasets consist of two components: benign queries and adversarial queries. Specifically, to calculate F1-score, we construct a balanced test set for each benchmark–method pair. Concretely, we select four benchmarks~(i.e., $D_{test}$, MMLU, GSM8K, MT-Bench) as the benign query datasets. Then, for every benign query in each dataset, we randomly sample one trigger from $\mathcal{T}{\text{test}}$, which includes 10 triggers for each attack method, and prepend it to the query to obtain the rerouting version. The overall statistics of the constructed test datasets are summarized in Table~\ref{table:dataset-num}.


\textbf{Baselines}. There are currently no specific guardrails against rerouting threats, thus we consider several common defense mechanisms against adversarial attacks, including perplexity-based filtering~(PPL)~\cite{alon2023detecting,jain2023baseline}, model-based filtering~\cite{metaPromptGuard,metallamaguard3}, and multi-router-based defense. The PPL method filters unnatural adversarial prompts with high perplexity, while ML-based filtering and LLM-based filtering use a deep learning model or LLM to identify and filter such unnatural prompts. The multiple router approach makes the router decision based on the majority vote of a set of LLM routers. 
Details of these defense methods are as follows: for PPL, following \cite{alon2023detecting,jain2023baseline}, we use GPT-2 to calculate the perplexity of user queries and set the PPL threshold to \textbf{223.59}, which corresponds to the highest perplexity observed among the benign queries in our selected dataset. This ensures that typical user queries will not trigger the detector.
For model-based filtering, we select the Prompt-Guard-86M~\cite{metaPromptGuard} and Llama-Guard-3-1B~\cite{metallamaguard3}, which are commonly used content safety guardrails. The Prompt-Guard-86M~\cite{metaPromptGuard} is a multi-category classifier that categorizes queries into three categories, i.e., benign, injection, and jailbreak prompts, where we consider injection and jailbreak prompts as adversarial. The Llama-Guard-3-1B~\cite{metallamaguard3} provides categorical safety judgments~(e.g., safe vs. unsafe and specific risk categories), rather than a continuous benignness score, and is used to exclude queries that fall into unsafe content categories.
For multiple-router defense, we use four LLM routers~(i.e., $R_{CLS}$, $R_{MF}$, $R_{SW}$, $R_{LLM}$) to jointly determine which model to route user queries to.

\textbf{Metrics}. 
We use detection accuracy and F1-score to evaluate the detection performance of filtering-based defenses, while the attack success rate~(ASR) is used to assess the mitigation performance of all defense methods. The detection accuracy reflects the ability of defenses to identify rerouting prompts. 
The F1-score, which incorporates precision and recall, provides insight into the false positive rate of detection methods, i.e., whether normal queries are mistakenly identified as rerouting prompts. The ASR, on the other hand, reflects the performance of defenses to mitigate rerouting risks.


\begin{table}[t!]
	\centering
	\caption{The statisitics of our test datasets. Each benchmark–method pair consists equal number of benign queries and rerouting queries.}
	\label{table:dataset-num}
	\begin{tabular}{lcccc}
		\toprule
		\textbf{Setting} & $D_{test}$ & \textbf{MMLU} & \textbf{GSM8k} & \textbf{MT-Bench} \\
		\cmidrule{1-5}
		White-box setting & 7712 & 200 & 200 & 144\\
		\midrule
		
		Gray-box setting & 7712 & 200 & 200 & 144\\
		\midrule
		
		Box-free setting & 7712 & 200 & 200 & 144\\
		\bottomrule
	\end{tabular}
	\vspace{-5pt}
\end{table}

\textbf{Environment}.
We conduct our experiments on a server equipped with 5 NVIDIA RTX 5880 GPUs. The server runs on the Ubuntu 20.04.6 LTS operating system. The experiments utilize Python 3.12.2, CUDA 12.8, PyTorch 2.8.0, and the Transformers library 4.57.0.

\begin{table*}[ht!]
	\centering
	\caption{Performance of different rerouting detection methods.}
	\label{tab:detection}
	\resizebox{.85\linewidth}{!}{
			\begin{tabular}{lcccc|cccc}
				\toprule
				\multirow{2}{*}{\textbf{Methods}} & \multicolumn{8}{c}{\textbf{Accuracy}$\uparrow$ / \textbf{F1-score}$\uparrow$} \\ \cmidrule{2-9}
				& \multicolumn{4}{c|}{$D_{test}$} & \multicolumn{4}{c}{MMLU} \\
				\midrule
				& Gray-box & Box-free & White-box & \textbf{Average} & Gray-box & Box-free & White-box & \textbf{Average} \\
				\midrule
				PPL  & 0.74/0.71 & 0.43/0.00 & 0.87/0.87 & 0.68/0.53 & 0.50/0.00 & 0.50/0.00 & 0.50/0.00 & 0.50/0.00 \\
				PromptGuard  & 0.49/0.65 & 0.50/0.67 & 0.45/0.61 & 0.48/0.64 & 0.50/0.67 & 0.50/0.67 & 0.50/0.67 & 0.50/0.67 \\
				LlamaGuard  & 0.52/0.52 & 0.49/0.50 & 0.52/0.52 & 0.51/0.52 & 0.44/0.46 & 0.54/0.53 & 0.54/0.54 & 0.50/0.51 \\
				\rowcolor[HTML]{e6e6e6}
				Ours  & 1.00/1.00 & 1.00/1.00 & 1.00/1.00 & 1.00/1.00 & 1.00/1.00 & 1.00/1.00 & 1.00/1.00 & 1.00/1.00 \\
				\midrule
				& \multicolumn{4}{c|}{GSM8k} & \multicolumn{4}{c}{MT-Bench} \\
				\midrule
				& Gray-box & Box-free & White-box & \textbf{Average} & Gray-box & Box-free & White-box & \textbf{Average} \\
				\midrule
				PPL  & 0.53/0.10 & 0.50/0.00 & 0.52/0.06 & 0.51/0.05 & 0.68/0.53 & 0.50/0.00 & 0.72/0.62 & 0.63/0.38 \\
				PromptGuard  & 0.50/0.67 & 0.50/0.67 & 0.50/0.67 & 0.50/0.67 & 0.47/0.64 & 0.50/0.67 & 0.44/0.62 & 0.47/0.64 \\
				LlamaGuard  & 0.55/0.52 & 0.49/0.46 & 0.51/0.51 & 0.52/0.50 & 0.48/0.45 & 0.58/0.57 & 0.47/0.47 & 0.51/0.50 \\
				\rowcolor[HTML]{e6e6e6}
				Ours  & 1.00/1.00 & 1.00/1.00 & 0.99/0.99 & 1.00/1.00 & 1.00/1.00 & 1.00/1.00 & 1.00/1.00 & 1.00/1.00 \\
				\bottomrule
			\end{tabular}
		}
	\end{table*}

	\subsection{Experimental Results}
	\label{sec:defense-results}
	
	\subsubsection{Detection Performance}
	\label{sec:defense-detection}
	We compare the performance of RerouteGuard in detecting rerouting prompts generated by three attack methods under benign queries from four benchmarks. Note that we consistently select an equal number of test normal prompts and test rerouting prompts to compute detection accuracy and F1-score. This ensures that detection methods perform well in identifying rerouting prompts and the false positive rate for normal samples is demonstrated.
	
	As illustrated in Table~\ref{tab:detection}, our method consistently achieves nearly perfect performance across all attack settings and benchmarks, with accuracy and F1-scores predominantly close to 1.00. This demonstrates the effectiveness and generalizability of our approach in detecting rerouting prompts under various threat models, regardless of the underlying task or domain. 
	In contrast, baseline methods exhibit significant limitations. 
	The PPL-based method shows moderate performance when detecting rerouting queries with unreadable adversarial triggers~(i.e., gray-box and white-box attack), but fails completely in identifying semantically meaningful rerouting prompts~(i.e., box-free attack), with a detection accuracy of 0\%. Besides, in some benchmarks, the queries with unreadable triggers show similar perplexity with normal queries~(Table~\ref{table:trigger_pattern}), which further limits the effectiveness of PPL.
	The PromptGuard method, while detecting rerouting prompts with an accuracy of more than 87\%, misclassifies nearly all normal prompts as malicious queries, resulting in a high false positive rate.
	The LlamaGuard achieves an accuracy of 0.5 for both normal and rerouting prompts, close to random guessing, which suggests that existing methods for detecting adversarial attacks cannot transfer to rerouting attack scenarios effectively.
	These results validate the effectiveness of our method in reliably identifying rerouting attempts across different scenarios. 

			\subsubsection{Mitigation Performance}
			\label{sec:defense-mitagation}
			
			We evaluate the performance of our method with four baselines in mitigating rerouting risks. Specifically, for filtering-based defenses, we drop out the specious queries, while for multi-router defense, we conduct a majority vote on their routing decisions. 
			We present the results on $D_{test}$ in Table~\ref{tab:mitigation}. 
			
			It could be observed that, without any defense deployed, all attack methods exhibit an ASR close to 1 on $D_{test}$, indicating that these simple queries can be easily manipulated to reroute to the strong model. 
			However, after deploying our defense, the ASR drops to 0 for all methods, showing that our approach can effectively mitigate rerouting attacks. 
			In comparison, other baselines show limited performance. For example, PPL struggles to detect semantically meaningful rerouting prompts~(i.e., box-free attacks), where the ASR remains close to 1 even after defense. On the other hand, LlamaGuard performs like random guessing, filtering out about half of the rerouting queries and resulting in an ASR of 0.5. Notably, after deploying PromptGuard, the ASR also drops close to 0, but it comes at the cost of a significant performance decline on normal queries, as shown in Section~\ref{sec:defense-mitagation}. Additionally, MultiRouter also has little to no effect, suggesting that these triggers are transferable across different routers and cannot be defended against through the joint decisions of multiple routers. 
			The experimental results demonstrate the superior performance of our method in mitigating rerouting risks.
			
			\begin{table}[t!]
				\centering
				\caption{Performance of different rerouting mitigation methods. The ``Avg'' means average ASR.}
				\label{tab:mitigation}
				\resizebox{\linewidth}{!}{
						\begin{tabular}{llcccc}
							\toprule
							\multirow{2}{*}{\textbf{Routers}} & \multirow{2}{*}{\textbf{Methods}} & \multicolumn{3}{c}{\textbf{Attack Success Rate}$\downarrow$} & \multirow{2}{*}{\textbf{Avg}$\downarrow$} \\
							\cmidrule{3-5}
							& & Gray-box & Box-free & White-box &\\
							\midrule
							\multirow{6}{*}{$R_{CLS}$} & No-defense & 1.00 & 1.00 & 1.00 & 1.00 \\
							& PPL & 0.34 & 1.00 & 0.06 & 0.47 \\
							& PromptGuard & 0.02 & 0.00 & 0.09 & 0.04 \\
							& LlamaGuard & 0.48 & 0.47 & 0.55 & 0.50 \\
							& Multi-Router & 0.99 & 1.00 & 1.00 & 1.00 \\
							\rowcolor[HTML]{e6e6e6}
							& Ours & 0.00 & 0.00 & 0.00 & 0.00 \\
							\midrule
							\multirow{6}{*}{$R_{MF}$} & No-defense & 0.92 & 0.98 & 0.88 & 0.93 \\
							& PPL & 0.33 & 0.98 & 0.09 & 0.47 \\
							& PromptGuard & 0.03 & 0.00 & 0.09 & 0.04 \\
							& LlamaGuard & 0.49 & 0.47 & 0.41 & 0.46 \\
							& Multi-Router & 0.99 & 1.00 & 1.00 & 1.00 \\
							\rowcolor[HTML]{e6e6e6}
							& Ours & 0.00 & 0.00 & 0.00 & 0.00 \\
							\midrule
							\multirow{6}{*}{$R_{SW}$} & No-defense & 0.96 & 1.00 & 0.99 & 0.98 \\
							& PPL & 0.34 & 1.00 & 0.07 & 0.47 \\
							& PromptGuard & 0.04 & 0.00 & 0.11 & 0.05 \\
							& LlamaGuard & 0.50 & 0.48 & 0.47 & 0.48 \\
							& Multi-Router & 0.99 & 1.00 & 1.00 & 1.00 \\
							\rowcolor[HTML]{e6e6e6}
							& Ours & 0.00 & 0.00 & 0.00 & 0.00 \\
							\midrule
							\multirow{6}{*}{$R_{LLM}$} & No-defense & 0.86 & 0.93 & 1.00 & 0.93 \\
							& PPL & 0.27 & 0.93 & 0.08 & 0.43 \\
							& PromptGuard & 0.03 & 0.00 & 0.10 & 0.04 \\
							& LlamaGuard & 0.45 & 0.44 & 0.49 & 0.46 \\
							& Multi-Router & 0.99 & 1.00 & 1.00 & 1.00 \\
							\rowcolor[HTML]{e6e6e6}
							& Ours & 0.00 & 0.00 & 0.00 & 0.00 \\
							\bottomrule
						\end{tabular}
					}
					\vspace{-10pt}
				\end{table}

				\subsubsection{Utility and Efficiency} In previous experiments, we have demonstrated that RerouteGuard rarely misclassifies normal queries. To further validate its minimal impact on normal performance, we investigate its capability in distinguishing truly complex queries from rerouting queries. Specifically, we employ a similar construction approach as $D_{normal}$ and sample the queries whose win rates are consistently within the top 50\% across all routers, resulting in 11,421 complex questions referred to as $D_{complex}$. The experimental results show that our method still achieves an accuracy of 0.96, which indicates that it effectively distinguishes between genuinely complex queries and malicious rerouting queries, thus having almost no impact on the performance of normal routers.
				
				In addition, RerouteGuard is lightweight and can obtain classification results in just 0.009 seconds when given a new query. Compared to the inference time of routers, where $R_{CLS}$ is 0.04s, $R_{LLM}$ is 0.21s, $R_{MF}$ is 1.66s and $R_{SW}$ is 8.00s, RerouteGuard is an efficient and easily integrable solution for real-time rerouting defense, without sacrificing performance.
				

				\begin{table}[t!]
					\centering
					\caption{Performance of RerouteGuard against adaptive attacks.}
					\label{tab:adaptive-attack}
						\scalebox{0.9}{
							\begin{tabular}{llccc}
								\toprule
								\multirow{2}{*}{\textbf{Routers}} & \multirow{2}{*}{\textbf{Methods}} & \multicolumn{2}{c}{\textbf{Attack Success Rate}$\downarrow$} & \multirow{2}{*}{\textbf{Avg}$\uparrow$} \\
								\cmidrule{3-4}
								& & Gray-box & White-box &\\
								\midrule
								\multirow{2}{*}{$R_{CLS}$} & No-defense & 1.00 & 0.99 & 1.00 \\
								& RerouteGuard & 0.00 & 0.32 & 0.16 \\
								\midrule
								\multirow{2}{*}{$R_{MF}$} & No-defense & 0.98 & 0.86 & 0.92 \\
								& RerouteGuard & 0.00 & 0.26 & 0.13 \\
								\midrule
								\multirow{2}{*}{$R_{SW}$} & No-defense & 0.98 & 0.93 & 0.96 \\
								& RerouteGuard & 0.00 & 0.27 & 0.14 \\
								\midrule
								\multirow{2}{*}{$R_{LLM}$} & No-defense & 0.80 & 0.98 & 0.89 \\
								& RerouteGuard & 0.00 & 0.31 & 0.15 \\
								\bottomrule
							\end{tabular}
						}
					\end{table}

					\subsubsection{Performance against Adaptive Attacks} To evaluate the robustness of RerouteGuard against adaptive attacks, we modify the optimization objetives for both the gray-box and white-box attack, considering both rerouting success rate and stealithness. For gray-box attack, we incorporate the adversarial probability predicted by RerouteGuard into the optimization feedback as a penalty term:
					\begin{equation}
						t_{m+1} = \max_{t \in N(t_m)} \big[P_{\theta}(M_{\mathrm{target}} \mid t ) - \alpha \cdot \text{RerouteGuard}(t)\big]
					\end{equation}
					where $\text{RerouteGuard}(t)$ denotes the adversarial probability predicted by RerouteGuard, and $\alpha$ is a hyperparameter, which we set to 0.5 in our experiments. 
					For white-box attack, we add the RerouteGuard feedback as an additional loss term, i.e., 
					\begin{equation}
						\mathcal{L}_{\text{defense}} = -\log(\text{RerouteGuard}(t \oplus q) + \epsilon)
					\end{equation}
					\begin{equation}
						\mathcal{L}_\text{combined} = \alpha * \mathcal{L}_\text{router} + \beta * \mathcal{L}_\text{defense} 
					\end{equation}
					where $q$ is the query from the white-box attack training dataset, $\epsilon$ is a small constant, $\alpha$ and $\beta$ are hyperparameters. We set $\epsilon=1e-8$, $\alpha=\beta=0.5$ in our exeriments.
					We do not consider the box-free attack as an adaptive attack, since it does not rely any feedback information to obtain triggers.
					
					We present the ASR results of RerouteGuard against adaptive attacks in Table~\ref{tab:adaptive-attack}. The results demonstrate that RerouteGuard remains highly robust even when confronted with adaptive rerouting queries of gray-box attack, dropping the ASR to 0\%. However, it shows relatively weaker performance in white-box adaptive attack, dropping the ASR to around 30\%. This is possibly due to the more informative optimization signal, which enables the attacker to exploit subtle weaknesses in the learned decision boundary of RerouteGuard, resulting in partially successful adaptive attacks.

					\subsection{Ablation Study}
					\label{sec:defense-ablation}
					
					The core of RerouteGuard is contrastive learning. To verify its necessity, we perform an ablation study. Specifically, we change the training procedure to directly train a binary classification detector, i.e., directly determining whether a query is a normal prompt or a rerouting prompt, instead of constructing pairs. 
					As shown in Table~\ref{tab:ablation-single-query}, the single-query classifier performs well on $D_{test}$ and MT-Bench~(open-ended questions), with an F1-score greater than 0.95. However, the F1-score drops to around 0.7 on MMLU and GSM8K because the single-query classifier misclassifies more than 85\% normal prompts as rerouting prompts. This indicates that without contrastive learning, the model may only learn single normal and attack patterns, resulting in poor generalization. Our method, however, captures more fine-grained discriminative information by learning the relative relationships between samples, thus enabling excellent transfer performance across different benchmarks.

					\begin{table}[t!]
						\centering
						\caption{The performance comparison between our defense and single query classfier~(i.e., without contrastive learning).}
						\label{tab:ablation-single-query}
						\resizebox{\linewidth}{!}{
							\begin{tabular}{llcccc}
								\toprule
								\multirow{2}{*}{\textbf{Benchmarks}} & \multirow{2}{*}{\textbf{Methods}} & \multicolumn{3}{c}{\textbf{Accuracy}$\uparrow$ / \textbf{F1-score}$\uparrow$} & \multirow{2}{*}{\textbf{Avg}$\uparrow$} \\
								\cmidrule{3-5}
								& & Gray-box & Box-free & White-box &\\
								\midrule
								\multirow{2}{*}{$D_{test}$}
								& w/o constr. & 1.00/1.00 & 1.00/1.00 & 1.00/1.00 & 1.00 \\
								& ours & 1.00/1.00 & 1.00/1.00 & 1.00/1.00 & 1.00 \\
								\midrule
								\multirow{2}{*}{MMLU}
								& w/o constr. & 0.57/0.70 & 0.57/0.70 & 0.57/0.70 & 0.70 \\
								& ours & 1.00/1.00 & 1.00/1.00 & 1.00/1.00 & 1.00 \\
								\midrule
								\multirow{2}{*}{GSM8k}
								& w/o constr. & 0.50/0.67 & 0.50/0.67 & 0.50/0.67 & 0.67 \\
								& ours & 1.00/1.00 & 1.00/1.00 & 0.99/0.99 & 1.00 \\
								\midrule
								\multirow{2}{*}{MT-Bench}
								& w/o constr. & 0.94/0.95 & 0.94/0.95 & 0.94/0.95 & 0.95 \\
								& ours & 1.00/1.00 & 1.00/1.00 & 1.00/1.00 & 1.00 \\
								\bottomrule
							\end{tabular}
						}
					\end{table}

\section{Conclusion}
In this work, we presented the first systematic measurement study of LLM rerouting attacks. We began by formally characterizing the potential safety and security risks posed by LLM rerouting, focusing on the adversary's objectives, capabilities, and knowledge.
We then evaluated the adversarial risks of four mainstream routers against three attack methods. We revealed that these routers are highly vulnerable to adversarial rerouting, with an average attack success rate (ASR) exceeding 80\%.
We further analyzed the effectiveness of these attacks by conducting transferability and interpretability analyses. Our findings indicated that rerouting triggers often exploit the complexity of queries, which allows them to manipulate the decision boundaries of LLM routers. Buliding upon this analysis, we proposed RerouteGuard, a novel, lightweight guardrail designed to detect and block LLM rerouting attacks in a black-box setting. It leveraged contrastive learning to disentangle adversarial patterns from normal query complexity patterns, effectively filtering rerouting triggers. Experimental results show that RerouteGuard can detect almost all rerouting attacks with minimal impact on normal queries, achieving a detection accuracy of 100\% and an average false positive rate of less than 4\%.

\bibliographystyle{ACM-Reference-Format}
\bibliography{refs}


\begin{thebibliography}{29}


\ifx \showCODEN    \undefined \def \showCODEN     #1{\unskip}     \fi
\ifx \showISBNx    \undefined \def \showISBNx     #1{\unskip}     \fi
\ifx \showISBNxiii \undefined \def \showISBNxiii  #1{\unskip}     \fi
\ifx \showISSN     \undefined \def \showISSN      #1{\unskip}     \fi
\ifx \showLCCN     \undefined \def \showLCCN      #1{\unskip}     \fi
\ifx \shownote     \undefined \def \shownote      #1{#1}          \fi
\ifx \showarticletitle \undefined \def \showarticletitle #1{#1}   \fi
\ifx \showURL      \undefined \def \showURL       {\relax}        \fi
\providecommand\bibfield[2]{#2}
\providecommand\bibinfo[2]{#2}
\providecommand\natexlab[1]{#1}
\providecommand\showeprint[2][]{arXiv:#2}

\bibitem[not(2025)]%
        {notdiamond}
 \bibinfo{year}{2025}\natexlab{}.
\newblock \bibinfo{booktitle}{\emph{Notdiamond LLM router}}.
\newblock
\urldef\tempurl%
\url{https://www.notdiamond.ai/}
\showURL{%
\tempurl}


\bibitem[oll(2025)]%
        {ollama}
 \bibinfo{year}{2025}\natexlab{}.
\newblock \bibinfo{booktitle}{\emph{Ollama}}.
\newblock
\urldef\tempurl%
\url{https://ollama.com/}
\showURL{%
\tempurl}


\bibitem[rep(2025)]%
        {replicated}
 \bibinfo{year}{2025}\natexlab{}.
\newblock \bibinfo{booktitle}{\emph{Replicated}}.
\newblock
\urldef\tempurl%
\url{https://www.replicated.com/}
\showURL{%
\tempurl}


\bibitem[Aggarwal et~al\mbox{.}(2024)]%
        {aggarwal2024automix}
\bibfield{author}{\bibinfo{person}{Pranjal Aggarwal}, \bibinfo{person}{Aman
  Madaan}, \bibinfo{person}{Ankit Anand}, \bibinfo{person}{Srividya~Pranavi
  Potharaju}, \bibinfo{person}{Swaroop Mishra}, \bibinfo{person}{Pei Zhou},
  \bibinfo{person}{Aditya Gupta}, \bibinfo{person}{Dheeraj Rajagopal},
  \bibinfo{person}{Karthik Kappaganthu}, \bibinfo{person}{Yiming Yang},
  {et~al\mbox{.}}} \bibinfo{year}{2024}\natexlab{}.
\newblock \showarticletitle{AutoMix: Automatically mixing language models}.
\newblock \bibinfo{journal}{\emph{Advances in Neural Information Processing
  Systems}}  \bibinfo{volume}{37} (\bibinfo{year}{2024}),
  \bibinfo{pages}{131000--131034}.
\newblock


\bibitem[AI(2025)]%
        {PROMISQROUTE}
\bibfield{author}{\bibinfo{person}{Adversa AI}.}
  \bibinfo{year}{2025}\natexlab{}.
\newblock \bibinfo{booktitle}{\emph{PROMISQROUTE: GPT-5 AI Router Novel
  Vulnerability Class Exposes the Fatal Flaw in Multi-Model Architectures}}.
\newblock
\urldef\tempurl%
\url{https://adversa.ai/blog/promisqroute-gpt-5-ai-router-novel-vulnerability-class/}
\showURL{%
\tempurl}
\newblock
\shownote{Accessed: 2025-10-03}.


\bibitem[AI(2019)]%
        {facebookai2019xlm-roberta-base}
\bibfield{author}{\bibinfo{person}{Facebook AI}.}
  \bibinfo{year}{2019}\natexlab{}.
\newblock \bibinfo{title}{XLM-RoBERTa-base}.
\newblock


\bibitem[Alon and Kamfonas(2023)]%
        {alon2023detecting}
\bibfield{author}{\bibinfo{person}{Gabriel Alon} {and} \bibinfo{person}{Michael
  Kamfonas}.} \bibinfo{year}{2023}\natexlab{}.
\newblock \showarticletitle{Detecting language model attacks with perplexity}.
\newblock \bibinfo{journal}{\emph{arXiv preprint arXiv:2308.14132}}
  (\bibinfo{year}{2023}).
\newblock


\bibitem[Chao et~al\mbox{.}(2023)]%
        {chao2023PAIR}
\bibfield{author}{\bibinfo{person}{Patrick Chao}, \bibinfo{person}{Alexander
  Robey}, \bibinfo{person}{Edgar Dobriban}, \bibinfo{person}{Hamed Hassani},
  \bibinfo{person}{George~J Pappas}, {and} \bibinfo{person}{Eric Wong}.}
  \bibinfo{year}{2023}\natexlab{}.
\newblock \showarticletitle{Jailbreaking black box large language models in
  twenty queries}.
\newblock \bibinfo{journal}{\emph{arXiv preprint arXiv:2310.08419}}
  (\bibinfo{year}{2023}).
\newblock


\bibitem[Cobbe et~al\mbox{.}(2021)]%
        {cobbe2021gsm8k}
\bibfield{author}{\bibinfo{person}{Karl Cobbe}, \bibinfo{person}{Vineet
  Kosaraju}, \bibinfo{person}{Mohammad Bavarian}, \bibinfo{person}{Mark Chen},
  \bibinfo{person}{Heewoo Jun}, \bibinfo{person}{Lukasz Kaiser},
  \bibinfo{person}{Matthias Plappert}, \bibinfo{person}{Jerry Tworek},
  \bibinfo{person}{Jacob Hilton}, \bibinfo{person}{Reiichiro Nakano},
  {et~al\mbox{.}}} \bibinfo{year}{2021}\natexlab{}.
\newblock \showarticletitle{Training verifiers to solve math word problems}.
\newblock \bibinfo{journal}{\emph{arXiv preprint arXiv:2110.14168}}
  (\bibinfo{year}{2021}).
\newblock


\bibitem[Ding et~al\mbox{.}(2024)]%
        {ding2024hybrid}
\bibfield{author}{\bibinfo{person}{Dujian Ding}, \bibinfo{person}{Ankur
  Mallick}, \bibinfo{person}{Chi Wang}, \bibinfo{person}{Robert Sim},
  \bibinfo{person}{Subhabrata Mukherjee}, \bibinfo{person}{Victor Ruhle},
  \bibinfo{person}{Laks~VS Lakshmanan}, {and} \bibinfo{person}{Ahmed~Hassan
  Awadallah}.} \bibinfo{year}{2024}\natexlab{}.
\newblock \showarticletitle{Hybrid llm: Cost-efficient and quality-aware query
  routing}.
\newblock \bibinfo{journal}{\emph{arXiv preprint arXiv:2404.14618}}
  (\bibinfo{year}{2024}).
\newblock


\bibitem[Hendrycks et~al\mbox{.}(2021)]%
        {Dan2021mmlu}
\bibfield{author}{\bibinfo{person}{Dan Hendrycks}, \bibinfo{person}{Collin
  Burns}, \bibinfo{person}{Steven Basart}, \bibinfo{person}{Andy Zou},
  \bibinfo{person}{Mantas Mazeika}, \bibinfo{person}{Dawn Song}, {and}
  \bibinfo{person}{Jacob Steinhardt}.} \bibinfo{year}{2021}\natexlab{}.
\newblock \showarticletitle{Measuring Massive Multitask Language
  Understanding}.
\newblock \bibinfo{journal}{\emph{Proceedings of the International Conference
  on Learning Representations (ICLR)}} (\bibinfo{year}{2021}).
\newblock


\bibitem[Jain et~al\mbox{.}(2023)]%
        {jain2023baseline}
\bibfield{author}{\bibinfo{person}{Neel Jain}, \bibinfo{person}{Avi
  Schwarzschild}, \bibinfo{person}{Yuxin Wen}, \bibinfo{person}{Gowthami
  Somepalli}, \bibinfo{person}{John Kirchenbauer}, \bibinfo{person}{Ping-yeh
  Chiang}, \bibinfo{person}{Micah Goldblum}, \bibinfo{person}{Aniruddha Saha},
  \bibinfo{person}{Jonas Geiping}, {and} \bibinfo{person}{Tom Goldstein}.}
  \bibinfo{year}{2023}\natexlab{}.
\newblock \showarticletitle{Baseline defenses for adversarial attacks against
  aligned language models}.
\newblock \bibinfo{journal}{\emph{arXiv preprint arXiv:2309.00614}}
  (\bibinfo{year}{2023}).
\newblock


\bibitem[Li et~al\mbox{.}(2023)]%
        {li2023deepinception}
\bibfield{author}{\bibinfo{person}{Xuan Li}, \bibinfo{person}{Zhanke Zhou},
  \bibinfo{person}{Jianing Zhu}, \bibinfo{person}{Jiangchao Yao},
  \bibinfo{person}{Tongliang Liu}, {and} \bibinfo{person}{Bo Han}.}
  \bibinfo{year}{2023}\natexlab{}.
\newblock \showarticletitle{Deepinception: Hypnotize large language model to be
  jailbreaker}.
\newblock \bibinfo{journal}{\emph{arXiv preprint arXiv:2311.03191}}
  (\bibinfo{year}{2023}).
\newblock


\bibitem[Lin et~al\mbox{.}(2025)]%
        {lin2025life}
\bibfield{author}{\bibinfo{person}{Qiqi Lin}, \bibinfo{person}{Xiaoyang Ji},
  \bibinfo{person}{Shengfang Zhai}, \bibinfo{person}{Qingni Shen},
  \bibinfo{person}{Zhi Zhang}, \bibinfo{person}{Yuejian Fang}, {and}
  \bibinfo{person}{Yansong Gao}.} \bibinfo{year}{2025}\natexlab{}.
\newblock \showarticletitle{Life-Cycle Routing Vulnerabilities of LLM Router}.
\newblock \bibinfo{journal}{\emph{arXiv preprint arXiv:2503.08704}}
  (\bibinfo{year}{2025}).
\newblock


\bibitem[Liu et~al\mbox{.}(2023)]%
        {liu2023autodan}
\bibfield{author}{\bibinfo{person}{Xiaogeng Liu}, \bibinfo{person}{Nan Xu},
  \bibinfo{person}{Muhao Chen}, {and} \bibinfo{person}{Chaowei Xiao}.}
  \bibinfo{year}{2023}\natexlab{}.
\newblock \showarticletitle{Autodan: Generating stealthy jailbreak prompts on
  aligned large language models}.
\newblock \bibinfo{journal}{\emph{arXiv preprint arXiv:2310.04451}}
  (\bibinfo{year}{2023}).
\newblock


\bibitem[Llama~Team(2024a)]%
        {metallamaguard3}
\bibfield{author}{\bibinfo{person}{AI~@~Meta Llama~Team}.}
  \bibinfo{year}{2024}\natexlab{a}.
\newblock \bibinfo{title}{The Llama 3 Family of Models}.
\newblock
  \bibinfo{howpublished}{\url{https://github.com/meta-llama/PurpleLlama/blob/main/Llama-Guard3/1B/MODEL_CARD.md}}.
\newblock


\bibitem[Llama~Team(2024b)]%
        {metaPromptGuard}
\bibfield{author}{\bibinfo{person}{AI~@~Meta Llama~Team}.}
  \bibinfo{year}{2024}\natexlab{b}.
\newblock \bibinfo{title}{Llama Prompt Guard}.
\newblock
  \bibinfo{howpublished}{\url{https://github.com/meta-llama/PurpleLlama/blob/main/Prompt-Guard/MODEL_CARD.md}}.
\newblock


\bibitem[Mazeika et~al\mbox{.}(2024)]%
        {mazeika2024harmbench}
\bibfield{author}{\bibinfo{person}{Mantas Mazeika}, \bibinfo{person}{Long
  Phan}, \bibinfo{person}{Xuwang Yin}, \bibinfo{person}{Andy Zou},
  \bibinfo{person}{Zifan Wang}, \bibinfo{person}{Norman Mu},
  \bibinfo{person}{Elham Sakhaee}, \bibinfo{person}{Nathaniel Li},
  \bibinfo{person}{Steven Basart}, \bibinfo{person}{Bo Li}, {et~al\mbox{.}}}
  \bibinfo{year}{2024}\natexlab{}.
\newblock \showarticletitle{Harmbench: A standardized evaluation framework for
  automated red teaming and robust refusal}.
\newblock \bibinfo{journal}{\emph{arXiv preprint arXiv:2402.04249}}
  (\bibinfo{year}{2024}).
\newblock


\bibitem[Mehrotra et~al\mbox{.}(2024)]%
        {mehrotra2024TAP}
\bibfield{author}{\bibinfo{person}{Anay Mehrotra}, \bibinfo{person}{Manolis
  Zampetakis}, \bibinfo{person}{Paul Kassianik}, \bibinfo{person}{Blaine
  Nelson}, \bibinfo{person}{Hyrum Anderson}, \bibinfo{person}{Yaron Singer},
  {and} \bibinfo{person}{Amin Karbasi}.} \bibinfo{year}{2024}\natexlab{}.
\newblock \showarticletitle{Tree of attacks: Jailbreaking black-box llms
  automatically}.
\newblock \bibinfo{journal}{\emph{Advances in Neural Information Processing
  Systems}}  \bibinfo{volume}{37} (\bibinfo{year}{2024}),
  \bibinfo{pages}{61065--61105}.
\newblock


\bibitem[Ong et~al\mbox{.}(2024)]%
        {ong2024routellm}
\bibfield{author}{\bibinfo{person}{Isaac Ong}, \bibinfo{person}{Amjad
  Almahairi}, \bibinfo{person}{Vincent Wu}, \bibinfo{person}{Wei-Lin Chiang},
  \bibinfo{person}{Tianhao Wu}, \bibinfo{person}{Joseph~E Gonzalez},
  \bibinfo{person}{M~Waleed Kadous}, {and} \bibinfo{person}{Ion Stoica}.}
  \bibinfo{year}{2024}\natexlab{}.
\newblock \showarticletitle{Routellm: Learning to route llms with preference
  data}.
\newblock \bibinfo{journal}{\emph{arXiv preprint arXiv:2406.18665}}
  (\bibinfo{year}{2024}).
\newblock


\bibitem[OpenAI(2025)]%
        {gpt5}
\bibfield{author}{\bibinfo{person}{OpenAI}.} \bibinfo{year}{2025}\natexlab{}.
\newblock \bibinfo{booktitle}{\emph{GPT-5}}.
\newblock
\urldef\tempurl%
\url{https://openai.com/index/gpt-5-system-card/}
\showURL{%
\tempurl}


\bibitem[Ram{\'\i}rez et~al\mbox{.}(2024a)]%
        {ramirez2024optimising}
\bibfield{author}{\bibinfo{person}{Guillem Ram{\'\i}rez},
  \bibinfo{person}{Alexandra Birch}, {and} \bibinfo{person}{Ivan Titov}.}
  \bibinfo{year}{2024}\natexlab{a}.
\newblock \showarticletitle{Optimising calls to large language models with
  uncertainty-based two-tier selection}.
\newblock \bibinfo{journal}{\emph{arXiv preprint arXiv:2405.02134}}
  (\bibinfo{year}{2024}).
\newblock


\bibitem[Ram{\'\i}rez et~al\mbox{.}(2024b)]%
        {ramirez2024cache}
\bibfield{author}{\bibinfo{person}{Guillem Ram{\'\i}rez},
  \bibinfo{person}{Matthias Lindemann}, \bibinfo{person}{Alexandra Birch},
  {and} \bibinfo{person}{Ivan Titov}.} \bibinfo{year}{2024}\natexlab{b}.
\newblock \showarticletitle{Cache \& Distil: Optimising API Calls to Large
  Language Models}. In \bibinfo{booktitle}{\emph{ACL (Findings)}}.
\newblock


\bibitem[Shafran et~al\mbox{.}(2025)]%
        {shafran2025rerouting}
\bibfield{author}{\bibinfo{person}{Avital Shafran}, \bibinfo{person}{Roei
  Schuster}, \bibinfo{person}{Thomas Ristenpart}, {and} \bibinfo{person}{Vitaly
  Shmatikov}.} \bibinfo{year}{2025}\natexlab{}.
\newblock \showarticletitle{Rerouting llm routers}.
\newblock \bibinfo{journal}{\emph{arXiv preprint arXiv:2501.01818}}
  (\bibinfo{year}{2025}).
\newblock


\bibitem[Zeng et~al\mbox{.}(2024)]%
        {zeng2024PAP}
\bibfield{author}{\bibinfo{person}{Yi Zeng}, \bibinfo{person}{Hongpeng Lin},
  \bibinfo{person}{Jingwen Zhang}, \bibinfo{person}{Diyi Yang},
  \bibinfo{person}{Ruoxi Jia}, {and} \bibinfo{person}{Weiyan Shi}.}
  \bibinfo{year}{2024}\natexlab{}.
\newblock \showarticletitle{How johnny can persuade llms to jailbreak them:
  Rethinking persuasion to challenge ai safety by humanizing llms}.
\newblock \bibinfo{journal}{\emph{arXiv preprint arXiv:2401.06373}}
  (\bibinfo{year}{2024}).
\newblock


\bibitem[Zhang et~al\mbox{.}({[n.\,d.]})]%
        {zhang2024ecoassistant}
\bibfield{author}{\bibinfo{person}{Jieyu Zhang}, \bibinfo{person}{Ranjay
  Krishna}, \bibinfo{person}{Ahmed~Hassan Awadallah}, {and}
  \bibinfo{person}{Chi Wang}.} \bibinfo{year}{[n.\,d.]}\natexlab{}.
\newblock \showarticletitle{EcoAssistant: Using LLM Assistants More Affordably
  and Accurately}. In \bibinfo{booktitle}{\emph{ICLR 2024 Workshop on Large
  Language Model (LLM) Agents}}.
\newblock


\bibitem[Zheng et~al\mbox{.}(2023a)]%
        {zheng2023mtbench}
\bibfield{author}{\bibinfo{person}{Lianmin Zheng}, \bibinfo{person}{Wei-Lin
  Chiang}, \bibinfo{person}{Ying Sheng}, \bibinfo{person}{Siyuan Zhuang},
  \bibinfo{person}{Zhanghao Wu}, \bibinfo{person}{Yonghao Zhuang},
  \bibinfo{person}{Zi Lin}, \bibinfo{person}{Zhuohan Li},
  \bibinfo{person}{Dacheng Li}, \bibinfo{person}{Eric Xing}, {et~al\mbox{.}}}
  \bibinfo{year}{2023}\natexlab{a}.
\newblock \showarticletitle{Judging llm-as-a-judge with mt-bench and chatbot
  arena}.
\newblock \bibinfo{journal}{\emph{Advances in neural information processing
  systems}}  \bibinfo{volume}{36} (\bibinfo{year}{2023}),
  \bibinfo{pages}{46595--46623}.
\newblock


\bibitem[Zheng et~al\mbox{.}(2023b)]%
        {zheng2023judging}
\bibfield{author}{\bibinfo{person}{Lianmin Zheng}, \bibinfo{person}{Wei-Lin
  Chiang}, \bibinfo{person}{Ying Sheng}, \bibinfo{person}{Siyuan Zhuang},
  \bibinfo{person}{Zhanghao Wu}, \bibinfo{person}{Yonghao Zhuang},
  \bibinfo{person}{Zi Lin}, \bibinfo{person}{Zhuohan Li},
  \bibinfo{person}{Dacheng Li}, \bibinfo{person}{Eric.~P Xing},
  \bibinfo{person}{Hao Zhang}, \bibinfo{person}{Joseph~E. Gonzalez}, {and}
  \bibinfo{person}{Ion Stoica}.} \bibinfo{year}{2023}\natexlab{b}.
\newblock \bibinfo{title}{Judging LLM-as-a-judge with MT-Bench and Chatbot
  Arena}.
\newblock
\showeprint[arxiv]{2306.05685}~[cs.CL]


\bibitem[Zou et~al\mbox{.}(2023)]%
        {zou2023GCG}
\bibfield{author}{\bibinfo{person}{Andy Zou}, \bibinfo{person}{Zifan Wang},
  \bibinfo{person}{Nicholas Carlini}, \bibinfo{person}{Milad Nasr},
  \bibinfo{person}{J~Zico Kolter}, {and} \bibinfo{person}{Matt Fredrikson}.}
  \bibinfo{year}{2023}\natexlab{}.
\newblock \showarticletitle{Universal and transferable adversarial attacks on
  aligned language models}.
\newblock \bibinfo{journal}{\emph{arXiv preprint arXiv:2307.15043}}
  (\bibinfo{year}{2023}).
\newblock


\end{thebibliography}

\end{sloppypar}
\end{document}